\documentclass[%
 reprint,
showkeys,
 amsmath,amssymb,
 aps,
]{revtex4-2}
\usepackage{subcaption} 
\usepackage{graphicx}
\usepackage{dcolumn}
\usepackage{bm}
\usepackage{verbatim}

\DeclareMathOperator*{\trace}{tr}
\newcommand{\<}{\langle}
\renewcommand{\>}{\rangle}
\newcommand{\R}{\mathbb{R}}
\newcommand{\C}{\mathbb{C}}

\newcommand{\nPOVM}{M}
\newcommand{\nEachPOVM}{m} 
\renewcommand{\vec}[1]{\boldsymbol{#1}} 
\newcommand{\mat}[1]{\vec{#1}}

\renewcommand{\P}{\mathbb{P}} 
\renewcommand{\H}{\mathbb{H}} 
\newcommand{\fidelity}{F}
\newcommand{\model}{\hat{\fidelity}}
\renewcommand{\C}{\mathcal{C}}  

\newcommand{\hatC}{\widehat{\C}} 
\newcommand{\D}{\mathcal{D}}  
\newcommand{\X}{\mathcal{X}}  
\newcommand{\Y}{\mathcal{Y}}  
\newcommand{\identity}{\mat{\mathbb{I}}} 
\let\rhoNoBold\rho 
\renewcommand{\rho}{\mat{\rhoNoBold}}
\let\sigmaNoBold\sigma 
\renewcommand{\sigma}{\mat{\sigmaNoBold}}

\newcommand{\adjoint}{\dagger} 
\newcommand{\pinv}{+}

\usepackage{booktabs}
\usepackage{multirow}
\usepackage{algorithm}
\usepackage{algpseudocode}

\begin{document}

\preprint{APS/123-QED}

\title{Uncertainty Quantification for Quantum Fidelity Estimation}

\author{K. Aditi}
\email{email: kuchibhotla.aditi@colorado.edu}

\author{Stephen Becker}%
\email{email: stephen.becker@colorado.edu}
\affiliation{%
University of Colorado, Boulder, CO, USA
}


\begin{abstract} 
Quantum fidelity estimation is a fundamental task in quantum information and computing to benchmark the quality of quantum states --- and hence of quantum processes --- on noisy quantum hardware. Point estimates of fidelity by existing methods generally lack the rigorous uncertainty quantification required for reliable preparation and verification of these tasks. While Bayesian and bootstrap frameworks are widely used for uncertainty quantification, they are sensitive to model misspecification or lack formal coverage guarantees in non-asymptotic regimes. This paper introduces a hybrid ``Bayesian-informed'' conformal prediction-based framework to provide distribution-free, rigorous prediction intervals for quantum state fidelity estimation which can be used, with care, to approximate confidence intervals with respect to the quantum randomness of Born's rule. The framework requires a point estimator, and we consider both classical point estimators as well as data-driven neural net estimators. We evaluate the robustness of the methodology across a diverse range of experimental challenges, including finite-sample noise, tomographic incompleteness, physical model error, and out-of-distribution data. Our results demonstrate that conformalized estimators maintain statistical validity where traditional methods fail, offering a scalable and reliable approach for the verification of modern quantum systems. The best performing variant of our methodology is based on a class-conditional conformal prediction, as this most closely approximates the conditional coverage desired for a true confidence interval.
\end{abstract}

\maketitle


\section{Introduction}\label{sec:intro} 
State fidelity estimation is an important task in quantum information and quantum computation, serving as a metric for the characterization and verification of quantum device outputs. By quantifying the closeness between an experimentally prepared state and an ideal target, fidelity serves as a critical benchmark for the performance of quantum gates and the validation of quantum algorithms \cite{wang2022quantum}. Traditionally, fidelity is estimated by first performing quantum state tomography (QST) to estimate the unknown quantum state, and then plug-in the state estimator into the fidelity formula. However, the computational and experimental costs of QST scale exponentially with system size, rendering it impractical for large-scale quantum devices. While compressed sensing-based QST \cite{Gross2009QuantumSensing, flammia2012quantum, Kalev2015QuantumProtocols} recovers states using fewer measurements, it relies on structural assumptions and is computationally demanding. For this reason, there is increasing interest in methods which estimate the fidelity directly; we discuss this literature in \S\ref{subsec:litreview}.


Since quantum measurements are stochastic and often incomplete, an estimate of fidelity is never perfectly accurate, thus it is natural to want not just a point estimate of the fidelity but also an estimate of the uncertainty. Historically, there are two general styles of uncertainty quantification (UQ) of a scalar estimate: either estimate the full posterior distribution (using Bayesian assumptions), or estimate a confidence interval in which the true value is likely to be with a fixed probability such as $\ge95\%$. The former approach can also generate intervals (known as credible intervals) but the validity relies on the validity of the assumed prior distribution and is also computationally costly. 
%
A well-constructed confidence interval must balance two primary criteria: validity/coverage, ensuring the interval contains the true outcome with a prescribed probability; and efficiency, ensuring the interval is sufficiently narrow to be informative without being overly conservative. Historically, constructing confidence intervals has relied on strong assumptions on distribution of data, asymptotic approximations, or the availability of accurate priors, which may not hold in complex or finite-sample regimes. Furthermore, these intervals are often estimated via bootstrap resampling which does not have theoretical guarantees.


A third style of UQ product is the prediction interval (PI), which is like a confidence interval but for the prediction 
conditioned on a new sample, and is taken over a larger uncertainty space. 
The recent growth of conformal prediction (CP) techniques has enabled rigorous creation of PI in many settings.  
CP is a distribution-free framework that provides finite-sample coverage guarantees under minimal assumptions, independent of the underlying predictive model or the amount of training data. 
The approach is agnostic to the actual prediction mechanism, so it is sometimes termed ``black-box'', and can be applied to models like neural networks. It has seen recent success in high-stakes applications like safety-critical robotics \cite{luo2024sample, lekeufack2024conformal} and  medical imaging diagnostics \cite{lu2022fair, angelopoulos2022image}.
The specific form of CP we use is 
split conformal prediction, a calibration-based variant that leverages a held-out validation set to formalize such guarantees. In this approach, the predictive model's errors on the validation set are mapped to a scalar metric known as a conformity score. By computing a specific empirical quantile of these scores, the framework dynamically scales the width of the prediction intervals to ensure that future test errors are strictly bounded with a user-defined probability.

This paper 
develops an efficient 
hybrid approach combining Bayesian assumptions with 
split conformal prediction to generate  prediction intervals, and then we discuss and use simulations to empirically validate using these intervals as confidence intervals for fidelity estimation. 
We propose a neural network-based state fidelity estimator that can be combined with our UQ framework 
to output a prediction interval where the true fidelity lies with a user-defined probability. This guaranteed coverage, along with the dynamically scaled interval lengths, helps an experimentalist make more informed downstream decisions without having to trust the results blindly. We show that our proposed method overcomes the drawbacks of traditional uncertainty quantification methods, which are demanding in their assumptions, computationally unscalable, or intractable. Furthermore, we show that by combining our method with a Bayesian-style analysis, we can train the neural network over a global space of quantum states, conformalize its predictions, and verify its performance under different experimental scenarios. Specifically, by partitioning the target space into discrete fidelity bins, we overcome the gap between global marginal guarantees and the state-conditional needs of an actual experiment. This allows us to maintain robust, localized target coverage even when the system is subjected to out-of-distribution physical noise.

\section{Background}\label{sec:background}
For notation, let $\H^d$ denote the set of $d \times d$ Hermitian matrices; for $q$ qubit systems, then $d=2^q$. We generally use uppercase Latin letters ($X$) for random variables and lowercase ($x$) for their realizations, and use bold to emphasize when a quantity (e.g., $\rho$ or $\mat{A}_k$) is a matrix or vector, and calligraphic script (e.g., $\D$) for sets. The conjugate transpose of $\mat{A}$ is denoted $\mat{A}^\adjoint$ while the pseudo-inverse is denoted $\mat{A}^\pinv$. 
The fidelity is $\fidelity$, estimates of fidelity are $\model$, and $\identity$ is the identity matrix of appropriate size. The Greek symbols $\sigma,\rho$ are quantum states, $\pi$ is a distribution, $\theta$ is a generic parameter, and $\alpha$, $\beta$, $\delta$ and $\eta$ are scalars.

\subsection{Fidelity and quantum measurements}
Fidelity is a widely used metric for the verification and characterization of quantum states. Fidelity between a pure reference state $\sigma$ and a target state $\rho$ is given by
$\fidelity(\rho,\sigma)=\trace(\rho\sigma) =:\langle \rho, \sigma\rangle$
and thus for
a fixed pure $\sigma$, the fidelity is a linear functional of $\rho$.
Our setup is that $\sigma$ is known exactly (pen-and-paper), while $\rho$ is a physical state.
Estimating the fidelity---whether directly or indirectly---requires performing measurements on identically prepared copies of $\rho$. These measurements are described by a positive operator-valued measurement (POVM), a collection of $m$ positive semidefinite operators $\{\mat{A}_\ell\}_{\ell=1}^\nEachPOVM$ such that $\mat{A}_\ell\in\H^d$,  
$\mat{A}_\ell\succeq \mat{0}$ and $\sum_\ell \mat{A}_{\ell} = \identity$. A single measurement (or ``shot'') of the state $\rho$ with the POVM $\{\mat{A}_\ell\}_{\ell=1}^\nEachPOVM$ is an outcome $\ell\in\{1,\ldots,\nEachPOVM\}$ where the probability of observing $\ell$ is $\trace(\mat{A}_\ell \rho )$.  Given $N$ measurements on copies of the state $\rho$, these outcomes thus follow a multinomial distribution. The measurement data collected is the set of empirical frequencies $\{ f_\ell \}_{\ell=1}^\nEachPOVM$ where $f_\ell$ records the number of times outcome $\ell$ was observed divided by $N$. 

Especially when $m$ is small (as in the binary case when $m=2$), multiple POVMs are commonly used, i.e., using $M$ different POVMs. For the rest of the paper, we will assume each POVM is binary (and in general follow the configuration detailed in \cite{j5gh-hmtw}) which will allow us to simplify notation, though our framework applies to any collection of POVMs. For $m=2$, since $f_2 = 1-f_1$, we can describe the outcome with a single number, and henceforth the subscript $f_k$ will denote the output of the $k^\text{th}$ POVM, and $\vec{f}\in\R^M$ denotes the vector containing these outputs. Specifically, we use a Pauli-derived set of POVMs.
%
%
The $q$-qubit Pauli matrices $\mat{W}_1, \dots \mat{W}_{d^2}$ are defined as the $q$-fold tensor product of the single qubit Pauli matrices $\mathcal{W} = \{I, \sigma_x, \sigma_y, \sigma_z \}\subset \H^2$, where 
\begin{equation}\label{eq:POVM}
\mat{W}_k = \bigotimes_{j=1}^n w_j^{(k)}\in \H^d    
\end{equation}
for $w_j^{(k)}\in\mathcal{W}$ and $d=2^q$.  There are $4^n = d^2$ such matrices.  Each Pauli matrix defines a binary POVM  $\{\frac{1}{2} (\identity \pm \mat{W}_k)\}$, and it is this collection of Pauli POVMs that generate frequency data $f_k$. 
As is standard, we exclude the $q$-qubit identity matrix since its resulting POVM $\{\identity,\mat{0}\}$ is uninformative, leaving $\nPOVM=d^2-1$ POVMs.

\subsection{Fidelity estimation}\label{subsec:litreview}
Several methods have been proposed for the purpose of fidelity estimation alone, all of which  circumvent  full state reconstruction. Protocols such as Quantum State Verification (QSV),  Direct Fidelity Estimation (DFE) \cite{flammia2011direct, PhysRevLett.107.210404}, the SWAP test \cite{PhysRevLett.87.167902, Cincio_2018} and the entanglement witness \cite{PhysRevA.72.022310, Bourennane_2004} estimate fidelity directly, offering more computationally efficient alternatives to QST. While QSV and entanglement witness are specialized for certain pure states like $W$-state and stabilizer states, DFE works for arbitrary pure states. The DFE estimator also gives an uncertainty range, though this relies on some approximations so it is more of theoretical rather than practical interest.
The recent method in \cite{Seshadri_2024_PRL,Seshadri_2024} allows experimentalists to select any measurement scheme and also provides confidence interval for fidelity but at the expense of solving a large optimization problem in the pre-processing phase.

Another line of work proposes quantum-classical approaches~\cite{chen2022variational,cerezo2020variational} based on Variational Quantum Algorithms (VQA). Specifically, the method in \cite{cerezo2020variational} is a hybrid algorithm for low-rank fidelity estimation. The method gives a lower and an upper bounds for fidelity using truncated fidelity which uses the rank-$r$ approximation of the state. They show that these bounds tighten monotonically in $r$. While this algorithm can provide  bounds with good empirical performance, there is no statistical guarantee. Other fidelity estimation frameworks utilize randomized measurement \cite{PhysRevA.99.052323, PhysRevLett.124.010504}, matrix product states \cite{PhysRevA.101.032321, PhysRevX.8.031012} and classical shadows \cite{huang2020predicting, PRXQuantum.2.010307}. 
Several of these works rely heavily on low-rank or other structural assumptions.


Machine learning models have also been used for fidelity estimation. In particular, \cite{PhysRevLett.127.130503} proposes the use of neural network (NN) to estimate fidelity where the input to the network are the measurement statistics obtained through synthetic data generation. The error performance of the neural network is analyzed, providing insights into the trade-off between statistical accuracy and the number of distinct measurement settings among other comparisons. The approach in \cite{PhysRevLett.132.190801} overcomes the drawbacks of tailoring measurement settings to specific reference states or error levels by utilizing a small, fixed set of Pauli measurements, enabling generalization to larger systems. The framework introduced in \cite{PhysRevLett.130.210601} employs a convolutional neural network (CNN) to quantify state similarity by mapping measurement data to a low-dimensional state representation and is designed for continuous families of quantum states, for instance those found in quantum optics. The work in \cite{vadali2024quantum} utilizes a NN to estimate the fidelity of a quantum circuit, where the model---once trained on hardware-specific measurement data---provides predictions for previously unseen quantum circuits. Alternatively, indirect fidelity estimation can be achieved by utilizing the ML techniques outlined in \cite{torlai2018neural, ahmed2021quantum, carleo2019machine, koutny2022neural, schmale2022efficient, krenn2023artificial} to perform QST, 
in an attempt to avoid computationally expensive solvers like iterative maximum likelihood estimation (iMLE), though since these are indirect methods that estimate the full $d\times d$ state, they would naturally be slower and require more training data compared to direct methods.

\subsection{Uncertainty quantification}
We review standard methods used to find error bars on estimates.
\subsubsection{Bayesian Inference and Credible Intervals}\label{subsec:Bayesian}
In the Bayesian framework for uncertainty quantification, any knowledge regarding the underlying physics model is encoded into a prior distribution, which is updated upon observing the data to yield a posterior distribution. For a given likelihood function, representing the probability of observing the data given the model parameters and a specified prior, the posterior distribution over the parameter space is determined via Bayes' theorem. More formally, let $\theta$ denote an unknown parameter of any dimensionality (e.g., the quantum state $\rho$, or its fidelity with respect to a fixed state $\sigma$), and assume a prior distribution with probability distribution function $\pi(\theta)$.
Write $\mathcal{L}(X \mid \theta)$ for the likelihood function, i.e., 
the conditional probability density 
of observing data $X$ given parameters $\theta$. Upon observing experimental data $X$, the posterior distribution $p(\theta \mid X)$ is given by Bayes' theorem \footnote{We allow $\pi$ to be continuous or discrete, and interpret the integral as the Riemann–Stieltjes integral in the latter case}:
\begin{equation}\label{eq:posterior}
p(\theta \mid X) = \frac{\mathcal{L}(X \mid \theta)\,\pi(\theta)}{\int \mathcal{L}(X \mid \theta')\,\pi(\theta')\, d\theta'}.
\end{equation}

This posterior characterizes the uncertainty in the model parameters and enables probabilistic predictions for subsequent observations. To quantify uncertainty, credible intervals are constructed from posterior percentiles that contain the desired probability mass. In practice, characterizing the posterior often requires stochastic sampling techniques, such as Markov Chain Monte Carlo (MCMC), which can render Bayesian inference computationally intensive for high-dimensional quantum systems. Another significant limitation of Bayesian uncertainty quantification is its sensitivity to model misspecification. When the underlying physical process deviates from the assumed prior or likelihood structure, the resulting posterior may fail to maintain the nominal coverage, often leading to overconfident or biased intervals. We give an example of this in \S\ref{sec:drawbacks}.

\subsubsection{Bootstrap Resampling and Approximate Confidence Intervals}\label{subsec:bootstrap}

The bootstrap~\cite{efron1979bootstrap} is a popular resampling-based method used to quantify the uncertainty of an estimator $\hat{\theta} = g(X_1, \dots, X_n)$ derived from data drawn from an unknown data-generating distribution $P$. The distribution $P$ is approximated by the empirical distribution $\hat{P}_n$, which assigns a probability mass of $1/n$ to each observed data point. By repeatedly resampling with replacement from $\hat{P}_n$, the bootstrap numerically simulates the sampling distribution of $\hat{\theta}$. This empirical sampling distribution enables the direct estimation of standard errors, statistical bias, and confidence intervals for complex or highly nonlinear functionals where exact analytical derivations are intractable. These intervals are not guaranteed to have correct coverage when used with finitely many samples though there are asymptotic results, e.g., \cite{chernozhukov2023high}.

We distinguish between parametric and non-parametric bootstrap approaches. The former generates resampled datasets from a fitted parametric model, while the latter resamples directly from the observed data without imposing distributional assumptions. For the construction of confidence intervals, the choice among percentile, reverse-percentile (basic) \cite{Kushary01052000}, and bias-corrected and accelerated (BCa) \cite{efron1994introduction} bootstrap intervals is typically guided by the finite-sample behavior of the estimator and the geometry of the bootstrap distribution, including skewness and bias. While the percentile method is computationally intuitive, it may exhibit conservative coverage in small samples. The reverse-percentile method corrects for bias by reflecting quantiles around the point estimate, whereas the BCa method further adjusts for skewness to provide higher asymptotic accuracy. In practice, the BCa method often yields the most reliable coverage, particularly under non-Gaussian or heteroscedastic error distributions but this requires using a bias-correction parameter $z_0$ and an acceleration parameter $a$ determined through jackknife estimates making BCa an expensive procedure. For our work, we choose to use the percentile method in \S\ref{sec:drawbacks} as it avoids any additional computational overhead.

Bootstrap techniques are a popular method for quantum estimation since they are fairly easy to use. For example, 
the work in \cite{Keith2018JointMeasurements} simultaneously employs QST and quantum detector tomography (QDT) to characterize both quantum states and the measurements with the help of a single high-fidelity state preparation and a limited set of high-fidelity unitary operations along with the low-fidelity counterparts. They use the parametric bootstrap resampling technique to calculate $95\%$ confidence intervals for observables like fidelity and expectation values of operators. The confidence intervals are determined using reverse-percentile and bias-correcting bootstrap since the Maximum Likelihood Estimate (MLE) is known to produce biased estimates due to its non-linear nature and the nature of SDPs \cite{Silva2017InvestigatingTomography}; an explicit example of an overconfident interval given by bootstrap approaches for QST is in \cite{Seshadri_2024}. 
Another drawback is that the method requires solving hundreds or thousands of MLE problems, which is computationally expensive in high dimensions.

\subsubsection{Conformal Prediction and Prediction Intervals}\label{subsec:intro_to_conformal}
Conformal prediction \cite{10.5555/645528.657641, 10.1007/3-540-36755-1_29, 10.5555/1062391, https://doi.org/10.1111/rssb.12021} is a relatively recent framework for constructing prediction intervals that are distribution-free, model-agnostic, and have finite-sample coverage guarantees. Crucially, these guarantees do not require knowledge of the data-generating distribution. Instead, conformal prediction relies only on the assumption that the data is independent and identically distributed (i.i.d.), though even this can be relaxed to a weaker concept known as exchangeability. 

The most widely used and computationally efficient variant is the split conformal (SC) prediction \cite{10.1007/3-540-36755-1_29, papadopoulos2008inductive}. Formally, the \emph{whole} training data is split into two disjoint subsets: a training set and a calibration set. The training set is used exclusively to fit the predictive model, while the calibration set is held out from training and used solely to compute conformal scores for each calibration example. A conformal score $s$ is a negatively oriented statistic that quantifies the degree of nonconformity, or disagreement, between the model’s prediction and the observed target value. For a new test point, the prediction set contains those candidate outcomes whose (non-) conformity score is no worse than observed for most of the calibration data. 


We now make this precise. Let $(X_i, Y_i)\in \X \times \Y$, $i=1, \dots ,n+1$ be exchangeable or i.i.d.\ feature and target variable pairs from a distribution 
on $\X \times \Y$. Let $\D_1$, $\D_2$ be the training set and the calibration set respectively, such that, $\vert \D_1 \vert = n_1$, $\vert \D_2 \vert = n_2$ and $n_1 + n_2 = n$. Let $(X_{n+1}, Y_{n+1})$ be a new test point for which we need to construct the prediction interval. The model $\model_{n_{1}}$ is trained on $\D_1$ and its performance is measured by the score function, which we take to be the absolute error function $s((x, y), \model_{n_{1}}) = \vert y - \model_{n_{1}}(x) \vert$. Scores are calculated on the calibration set giving $S_i = \vert Y_i - \model_{n_{1}}(X_i)\vert, i \in \D_2$. For a user-defined miscoverage level (significance level) $\alpha$, let $\hat{q}_{n_{2}}$ be the $(1- \alpha)$-quantile of the scores $S_i$. 
Then a prediction set defined as
\begin{equation} \label{eq:predictionset}
    \hat C_n(x) = \left[ \model_{n_{1}}(x)-\hat q_{n_{2}}, \model_{n_{1}}(x)+\hat q_{n_{2}} \right]
\end{equation}
satisfies the guarantee
\begin{equation}\label{conformal-guarantee}
    1-\alpha
    \le
    \mathbb{P}\!\left(
    Y_{n+1}\in \hat C_n(X_{n+1})
    \right)
    \le
    1-\alpha+\frac{1}{n_2+1},
\end{equation}
where the lower bound coverage always holds and the upper bound holds when scores are absolutely distinct. 
The probability is taken over the randomness in the training, calibration, and the new test data.

The coverage and the average length of the prediction intervals are two metrics that quantify the performance of a UQ framework. Since the conformal framework is model-agnostic, the coverage is guaranteed regardless of how good or bad the model is at prediction. However, the prediction power of the model does have an effect on how small or wide the prediction intervals can be. Intuitively, with a good prediction model, CP can generate smaller prediction intervals, 
hence we still care about the predictive performance of the model itself. 

\subsection{Variants of Conformal Prediction} \label{subsec:variants}
The split conformal method treats all calibration points equally, without explicitly accounting for heteroscedasticity in the data, where the noise level varies across different regions of the feature space. Consequently, the resulting prediction intervals are globally uniform in length and may be too wide for \emph{easy} inputs or too narrow for \emph{hard} inputs. This limitation motivates the development of locally adaptive conformal methods, 
which adjust the interval length based on the predicted uncertainty, thereby producing intervals that better reflect the local structure of the data. There are several ways to achieve this. 

\subsubsection{Adaptive split conformal} \label{subsubsec:adaptivesplit}
This is a locally adaptive version of split conformal \cite{papadopoulos2011reliable, lei2018distribution}, where, in addition to the point estimator for mean, a second model $\hat r(x)$ is trained 
using the original input features but using the residuals from the prediction of the first model as targets/labels, that is, $\{(X_i,S_i), i \in \D_1\}$.
Once trained, the purpose of this network is to predict residuals on the calibration set and the test data. The residuals on the calibration set are used to scale the original score to incorporate the variance in the data as
\begin{equation}
    \tilde S_i = \frac{S_i}{\hat r(X_i)}
\end{equation}
This allows the conformal interval to scale adaptively: points where the model tends to have large residuals get wider intervals, and points where the model is confident get narrower intervals. We include this adaptive version for comparison purposes in \S\ref{sec:results} and refer to it as adaptive (local) NN.

\subsubsection{Quantile-based split conformal} \label{subsubsec:quantilesplit}

A direct approach to constructing prediction intervals is to estimate the relevant conditional quantiles. A quantile regression model aims to learn the conditional $\alpha$-quantile of $Y$ given $X=x$
\begin{equation*}
    q_\alpha(x)
    :=
    \inf \left\{
        y \in \mathbb{R} :
        P(Y \leq y \mid X=x) \geq \alpha
    \right\}.
\end{equation*}
In practice, the true conditional distribution of $Y$ given $X$ is unknown, and hence quantile regression only provides an estimate $\hat q_\alpha(x)$. For example, these estimates may be obtained using a neural network trained with the pinball loss. Prediction intervals constructed directly from the estimated lower and upper quantiles are therefore not guaranteed to satisfy Eq.~\eqref{conformal-guarantee}. Conformalized quantile regression (CQR) addresses this issue by combining the ideas of conformal prediction with classical quantile regression, thereby recovering a finite-sample marginal coverage guarantee under exchangeability. Further details on quantile regression and its conformalization can be found in \cite{romano2019conformalized} and references therein.

In particular, a quantile regression model is trained on $\D_1$ to fit two conditional quantile functions $\hat q_{\alpha_{\text{low}}}$ and $\hat q_{\alpha_{\text{high}}}$ using the pinball loss. The conformity scores are calculated on the calibration set $\D_2$ as: $S_i = \max \{\hat q_{\alpha_{\text{low}}}(X_i) - Y_i, Y_i - \hat q_{\alpha_{\text{high}}}(X_i)\}, i\in \D_2$. The prediction interval is then defined as:
\begin{align}
    \hat C (X_{n+1}) = & \Big[\hat q_{\alpha_\text{low}}(X_{n+1}) - Q_{1-\alpha}(S, \D_2), \\
    & \hat q_{\alpha_\text{high}}(X_{n+1}) + Q_{1-\alpha}(S, \D_2)\Big], \notag
\end{align}
where $Q_{1-\alpha}(S, \D_2)$ is the $(1-\alpha)(1+\frac{1}{n_2})$ quantile of $S := \{S_i \ \vert \ i \in \D_2\}$.

Due to its proven effectiveness \cite{romano2019conformalized}, we use CQR as one of the evaluation methods for comparisons in \S\ref{sec:results}. Specifically, we use asymmetric CQR which calibrates the lower and upper quantile bounds separately giving stronger coverage guarantee.

\subsubsection{Split conformal for conditional coverage} \label{subsubsec:conditional}
Standard SC provides marginal coverage, which means that for a random test data, the prediction interval contains the true value with a probability of $1-\alpha$. 
However, often conditional coverage is more useful, and SC needs to be modified in order to be used.
A type of conditional coverage called $X$-conditional is defined for when data $X=x$ and is a stronger notion of coverage: 
\begin{equation}
    \P\left(
Y_{n+1} \in \hat C_n(X_{n+1})
\mid X_{n+1} = x
\right) \ge
1-\alpha, \quad \forall x.
\end{equation}
It was shown in \cite{vovk2012conditional, 10.1111/rssb.12021} that $X$-conditional coverage is impossible to achieve without further assumptions on the data distribution. For this reason, several works define a relaxed version of conditional coverage known as approximate $X$-conditional coverage or class-conditional coverage; see \cite{NEURIPS2020_244edd7e,foygel2021limits, pmlr-v162-angelopoulos22a, guan2023localized, gibbs2025conformal}. 
In this model, one partitions $\X$ into a finite set of
class $\mathcal{G}_1,\ldots,\mathcal{G}_g$, and 
class-conditional coverage is defined as
\begin{equation}
    \P\left(
Y_{n+1} \in \hat C_n(X_{n+1})
\mid X_{n+1} \in \mathcal{G}_i
\right) \ge
1-\alpha, \; \forall i=1,\ldots,g.
\end{equation}
This framework partitions the calibration dataset according to the observed variable $X$ and independently applies the conformal prediction procedure to each partition. Such an approach ensures that the prescribed coverage probability is maintained locally within each class, preventing any systematic over- or under-coverage for specific classes, similar to stratified sampling methods. Another type of conditional coverage is class-conditional (or $Y$-conditioned) coverage \cite{ding2023class, ShiEtAl_ClassConditional, article, hechtlinger2018cautious, sadinle2019least}, defined as
\begin{equation}
    \P\left(
Y_{n+1} \in \hat C_n(X_{n+1})
\mid Y_{n+1} \in \mathcal{G}_i 
\right) \ge
1-\alpha, \; \forall i=1,\ldots,g, 
\end{equation}
where $\mathcal{G}_i$ now partitions $Y$ instead. For our purpose, we use the $Y$-conditioned variant, deferring a detailed explanation of its application to \S\ref{subsec:cond-cov}.

\subsubsection{Conformalized 
Bayesian prediction}
Our proposed methodology described in \S\ref{sec:methodology} can be seen as a conformalized version of Bayesian prediction. This particular method is adapted to the circumstances of fidelity estimation, but we note that we are not the first to combine CP with Bayesian methods, and give a brief overview of prior work below.

Bayesian credible intervals do not, in general, provide finite-sample frequentist coverage guarantees, and their coverage can deteriorate further under model misspecification. In particular, assumptions on dimensionality imposed by the Bayesian model may not accurately reflect the data-generating distribution \cite{fontana2023conformal}. To address this limitation, \cite{fong2021conformal} proposed conformalized Bayesian inference, building on earlier work that uses the Bayesian posterior predictive distribution to construct conformity scores \cite{Melluish2001,Vovk2005,Wasserman2011,Burnaev2014}. The resulting conformal procedure combines the predictive information supplied by the Bayesian model with the finite-sample marginal coverage guarantee of conformal prediction. A direct implementation of full conformal prediction would require updating the Bayesian posterior for each candidate test response, which can be computationally expensive when posterior inference relies on MCMC. An efficient importance-sampling approach developed in \cite{fong2021conformal} reuses posterior samples, thereby avoiding a separate MCMC run for every candidate response. Alternative Bayesian conformity scores have also been considered in \cite{deliu2025interplay}.

\section{Proposed UQ methodology for fidelity estimation} \label{sec:methodology}
To put our fidelity estimation setup in the machine learning / conformal prediction notation, an ``instance'' is defined by a realization of a quantum state $\rho\in\H^d$, and the corresponding 
 input feature vector $X\in\X=\R^M$ is the set of empirical frequencies $\vec{f}=\{f_k\}_{k=1}^M$ collected across the Pauli POVMs. The corresponding target $Y\in\Y=[0,1]$ is the true quantum state fidelity, calculated between the state $\rho$ and the pure reference state $\sigma\in\H^d$.
We will eventually make use of a model $\model_{n_{1}}$ trained on $\D_1$ that predicts fidelity given data $X$; for example, one can train a neural net (see \S\ref{subsec:NN}) or use QST/MLE methods to estimate $\hat{\rho}$ from $X$ and then return $\model(X) = \fidelity(\sigma,\hat{\rho})$. For the latter case, there is no training phase and the estimate is independent of $\D_1$.

\subsection{Bayesian-Informed Conformal Prediction} \label{subsec:BayesianCP}

From the perspective of an experimental quantum physicist, the primary objective is to obtain a state-conditional coverage guarantee for a specific, fixed underlying state $\rho$:
\begin{equation} \label{eq:state-conditional}
\mathbb{P}_{\scalebox{0.5}{$
\begin{aligned}
X_{n+1} &\sim \mathcal{L}(X \mid \rho) \\ 
Y_{n+1} &\sim \delta_{\langle \sigma,\rho \rangle}
\end{aligned}$
}} 
\left(
Y_{n+1} \in \hat C_n(X_{n+1}) \bigm| \rho
\right) \ge 1-\alpha,
\end{equation}
where for a given state $\rho$, the measurement data $X$ is random due to the likelihood defined by Born's rule, and the true fidelity $Y$ is deterministic for the fixed reference state $\sigma$ (hence encoded as a Dirac delta function). 
We refer to a set $\hat C_n(X_{n+1})$ having the above properties as a true \emph{confidence interval}.
However, exact distribution-free conditional guarantees are mathematically impossible to satisfy for all individual states $\rho$ in finite samples. 

Instead, the marginal conformal prediction framework guarantees validity on average over a joint data-generating distribution $P$:
\begin{equation} \label{eq:actualGuarantee}
\mathbb{P}_{\scalebox{0.5}{$
\begin{aligned}
\D_1 &\overset{\text{i.i.d.}}{\sim} P \\ 
\D_2 &\overset{\text{i.i.d.}}{\sim} P \\
(X_{n+1}, Y_{n+1}) &\overset{\text{i.i.d.}}{\sim} P
\end{aligned}$
}} 
\left(
Y_{n+1} \in \hat C_n(X_{n+1}) \right) \ge 1-\alpha, 
\end{equation}
where the joint distribution $P(\rho, X, Y)$ is given by:
\begin{align}
    P(\rho, X, Y) &= \pi(\rho) P(X \mid \rho) P(Y\mid \rho, X)  \\
    &= \pi(\rho) P(X \mid \rho) P(Y\mid \rho) \notag \\
    &= \pi(\rho) \mathcal{L}(X \mid \rho) \delta(Y-\langle \sigma,\rho \rangle). \notag
\end{align}
Here, $\pi$ represents a prior distribution over the unknown state $\rho$ (e.g., the Hilbert-Schmidt measure).
We refer to a set $\hat C_n(X_{n+1})$ satisfying Eq.~\eqref{eq:actualGuarantee} as a \emph{prediction interval}.


Thus, while the physicist desires a guarantee for a specific state $\rho$, standard CP instead guarantees that the prediction intervals will be valid over the ensemble of states drawn from $\pi$. In order to achieve state-conditioned coverage, we implement a hybrid strategy that wraps these Bayesian priors within the conformal framework, allowing us to investigate and optimize the state-conditional adaptivity of our estimator. This hybrid strategy is summarized in Algorithm~\ref{alg:cap} and is written assuming binary POVMs for conciseness but is easily extended to general POVMs.


\begin{algorithm}[H] 
\caption{Bayesian-Informed Conformal Prediction}\label{alg:cap}
\begin{algorithmic}[1]
\Require Reference state $\sigma\in\H^d$, collection of $M$ POVMs, prior distribution $\pi$, number of shots $N$ per POVM, number of training $n_1$ and calibration $n_2$ points.
\State $\D \gets \varnothing$
\For{$i = 1, \ldots, n_1+n_2$} \Comment{Build data sets $\D_1,\D_2$}
    \State Sample $\rho \sim \pi$ \Comment{e.g., Hilbert-Schmidt measure}
    \State Set $Y=\fidelity(\sigma,\rho)\in[0,1]$
    \State Sample $X=\vec{f} \sim \mathcal{L}(X \mid \rho)$ as follows: 
    \For{$k=1,\ldots,M$} \Comment{Assuming binary POVMs}
        \State $p_k = \langle \mat{W}_k, \rho \rangle$ \Comment{$\mat{W}_k$ as in Eq.~\eqref{eq:POVM}}
        \State Sample $f_k \sim B(N,\frac12(1+p_k))$ \Comment{Binomial distr.}
        \State Normalize $f_k \gets f_k / N$
    \EndFor
    \State Append $X,Y$ to $\D$
    \EndFor
    \State Partition $\D=\D_1 \mathbin{\mathaccent\cdot\cup} \D_2$ \Comment{$|\D_1|=n_1$, $|\D_2|=n_2$}
\State Train model $\model_{n_{1}}$ on $\D_1$ \Comment{e.g., neural net}
\State Compute scores on calibration data $\D_2$ using $\model_{n_{1}}$
\State Compute quantiles on calibration data $\D_2$ using scores
\State Create prediction intervals $\hat C_n$ using Eq.~\eqref{eq:predictionset}
\end{algorithmic}
\end{algorithm}


Properly applied, this procedure rigorously ensures Eq.~\eqref{eq:actualGuarantee} holds, and works for any desired distribution $\pi$ since CP is distribution-free. Unlike Bayesian methods, $\pi$ need not have any nice properties, since in CP there is no posterior sampling.  However, if one desires a guarantee like Eq.~\eqref{eq:state-conditional}, then one should choose $\pi$ using as much physics knowledge as possible, with the understanding that, as in Bayesian methods, the prediction interval is not the same as a confidence interval.  If one knew, for example, that $\rho$ was a noisy version of a target state (e.g., with a random strength of depolarizing noise), this would be used as the prior. We do not pursue such targeted priors in our synthetic experiments since using a strong prior in synthetic data generation and in data processing would lead to unrealistically optimistic results.

In order to test how well Eq.~\eqref{eq:actualGuarantee} relates to Eq.~\eqref{eq:state-conditional}, as well as report meaningful statistics, we use a modified version of Eq.~\eqref{eq:actualGuarantee} where the training and calibration data are sampled from $P$ using the prior $\pi$, but the testing data are sampled $(X_{n+1}, Y_{n+1}) \overset{\text{i.i.d.}}{\sim} P'$ where $P'$ is defined as $P$ but with a modified prior $\pi'$. We give empirical evidence from synthetic simulations that our coverage guarantee is fairly robust to this ``distribution shift,'' and much more robust than a purely Bayesian method.

\subsubsection{Experimental use}
While our results are all synthetic for the purposes of validation, the framework is easily applied to experimental data as well.  The reference state $\sigma$ is known, and the set of POVMs is also known, so one chooses a prior $\pi$ and then synthetically generates the training and calibration datasets $\D_1,\D_2$. These are cheap to generate, so it is easy to make $n_1$ and $n_2$ large. The procedure then creates the model $\model_{n_1}$ and prediction interval $\hat C_n$ which can be applied to the experimental data. We have made our source code available at \cite{RigorousMLE}.

\subsection{Fidelity estimation models} \label{subsec:models}
The model $\model_{n_1}$ can be any of the methods described in Sec.~\ref{subsec:litreview}. Here we discuss in more detail the methods we will use in the synthetic experiments.

\subsubsection{Plugin estimators} \label{subsec:plugin}
``Plugin'' estimators do state estimation, estimating $\hat{\rho}$ from the instance, and then returning $\model(X) = \fidelity(\sigma,\hat{\rho})$. While it is possible for these methods to need a training phase, the particular classical estimators we use do not make use of the training data $\D_1$. We specifically compare with:
\begin{itemize}
    \item[\texttt{PLS}] Projected least-squares (PLS)  estimates $\hat{\rho}$ by first solving the linear least-squares problem $\min_{\rho} \sum_{i=1}^M \left(
    \< \mat{W}_k, \rho \> - \tilde{p}_k \right)^2
    $ where $\tilde{p}_k = 2 f_k - 1$, which can be solved efficiently using standard numerical linear algebra; if $M < d^2$ or the measurements are otherwise not full-rank then the least-norm solution consistent with the measurements is used. The constraint $\rho=\rho^\adjoint$ can optionally be enforced also via linear algebraic methods, but the constraint $\rho \succeq 0$ cannot be, so the second step is to project the least squares solution onto the set  $\rho \succeq 0$ and $\trace(\rho)=1$ either via a true orthogonal projection (the approach we take in our simulations) or via rescaling. This method is simple and goes back many years, with some recent error analysis in \cite{Guta2018FastBounds}.
    
    \item[\texttt{MLE}] Maximum likelihood estimation (MLE) is similar to PLS but uses a convex non-quadratic objective function based on the likelihood, and explicitly enforces the physical constraints $\rho=\rho^\adjoint, \rho\succeq 0, \trace(\rho)=1$ during the optimization rather than using the sub-optimal (yet fast) two-step approach as in PLS. For further details on the objective and the software, please see \cite{j5gh-hmtw}.
\end{itemize}

\subsubsection{Neural Net Architecture and Training} \label{subsec:NN}
Because our setup allows one to generate nearly unlimited training data at low-cost, 
we naturally pursue a data-driven approach that directly learns the fidelity (with respect to a fixed $\sigma$) as a function of samples, without reconstructing an intermediate estimate of the state $\hat{\rho}$.  In particular,  we employ a shallow feed-forward neural network, with a single hidden layer of 256 neurons, which maps samples $X\in\X=\R^M$ to estimates of fidelity $Y\in\Y=[0,1]$ and is trained on the data in $\D_1$ using the squared loss function;
for the version used in \S\ref{subsubsec:quantilesplit}, the same architecture is used but the output is now in $\R^2$ and training is done with the pinball loss.
We found that this architecture is highly robust across different system sizes; the fixed hidden layer width remains effective as the number of qubits grows, despite the expanding input feature space. This consistency indicates that the network successfully captures the linear structure of the fidelity, meaning the model complexity does not require proportional scaling with the exponentially large Hilbert space --- at least within the regime investigated here.

We generate synthetic dataset for training and validation using a simulated measurement framework described in Algorithm~\ref{alg:cap}. The training set is constructed by sampling density matrices $\rho$ according to the Hilbert-Schmidt measure. For each sampled state, we simulate the measurement procedure described above to generate empirical frequencies, which serve as the input features $X$. The corresponding regression targets $Y$ are the fidelity values $\fidelity(\sigma,\rho)$ calculated with respect to a fixed, randomly selected pure reference state $\sigma$. This supervised learning framework enables the network to directly learn the mapping from statistical measurement outcomes to the state fidelity.

The architectural validity of our model is demonstrated in Fig.~\ref{fig:nn-loss} (see Appendix \ref{appendix}), which illustrates the consistent convergence of training and validation loss. We notice that the same network produces smaller error as the system size increases from $2$ qubits to $6$. This may be due to the fact that the network receives more redundant measurement constraints on the same target value and uses them to suppress statistical noise through averaging.

\section{Drawbacks of classical UQ methods} \label{sec:drawbacks}

The two other natural UQ methods are credible intervals via Bayesian methods, and confidence intervals via the bootstrap. While these methods have strengths, they also have disadvantages which we discuss.


\begin{figure*}[ht]
    \centering

    \begin{subfigure}[t]{\textwidth}
    \centering
        \includegraphics[width=0.45\textwidth, trim=0.2cm 0.2cm 0.2cm 0.2cm, clip]{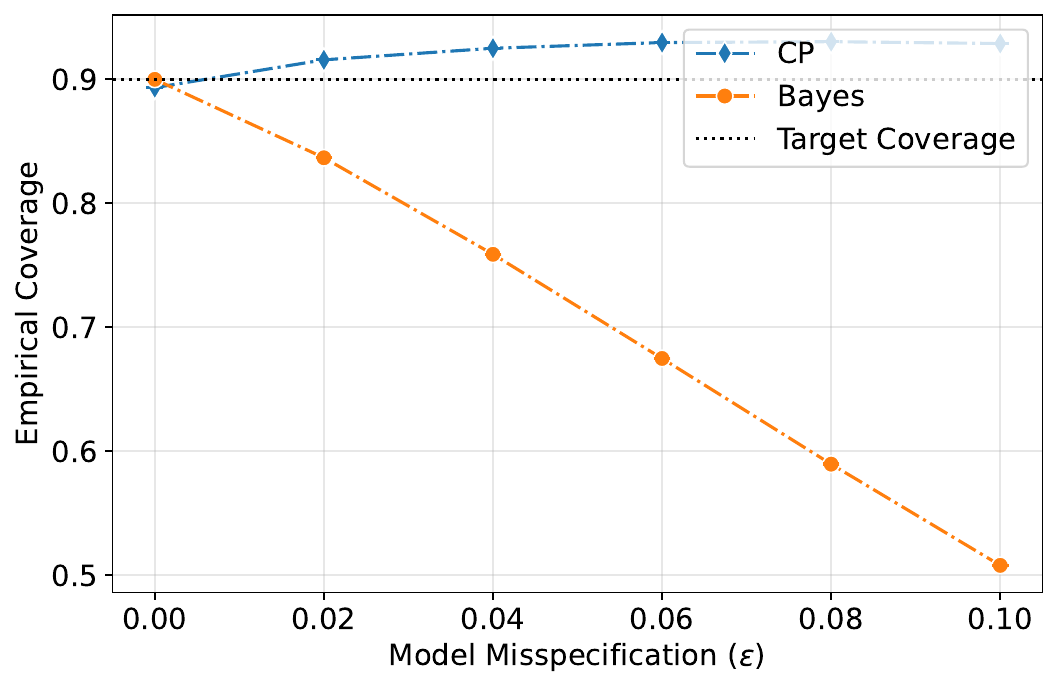}
        \hfill
        \includegraphics[width=0.45\textwidth, trim=0.2cm 0.2cm 0.2cm 0.2cm, clip]{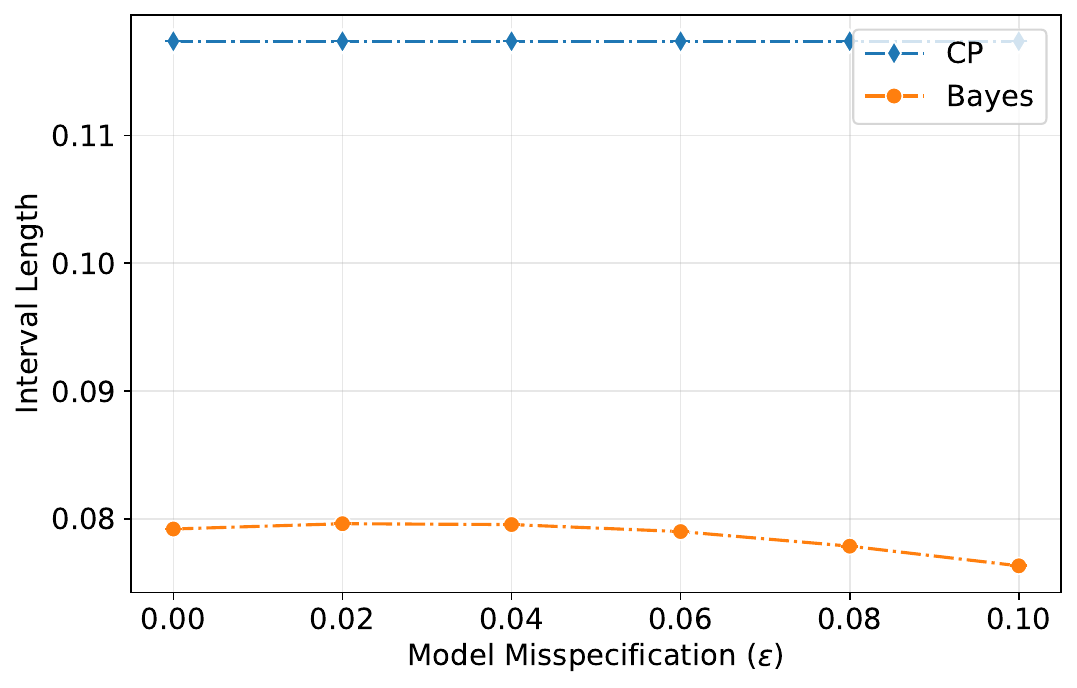}
    \end{subfigure}
    
    \caption{Impact of model misspecification on Bayesian prediction intervals. Empirical coverage (left) and interval length (right) as a function of model misspecification factor $\varepsilon$, compared against standard neural network with conformal prediction.}
    \label{fig:bayes}
\end{figure*}

\subsection{Use of incorrect prior in Bayesian framework} Below, we provide numerical simulations to characterize the impact of misspecification on coverage probability, highlighting the comparative robustness of our hybrid conformal prediction framework. 
General Bayesian methods require MCMC sampling which must be done right and is computationally expensive, so to avoid this distraction, we consider a  single parameter model that is simple enough to allow a closed-form posterior.
We restrict $\rho$ to be on a $1$-D line between two fixed states $\rho_1$ and $\rho_2$ which are chosen
as $\rho_1 = (1-\delta)\sigma + \delta \ \rho^\prime$ for an arbitrary state $\rho'$, and $\rho_2 = \sigma$.
Then our Bayesian prior $\pi$ is
\[
\rho = (1-\theta) \rho_1 + \theta \rho_2, \quad \theta \sim \text{Uniform}[0,1]. \\
\]

However, the actual (``physical'') states are drawn from a shifted distribution,
\begin{align*}
\rho_\text{phy} &= (1-\varepsilon)\rho + \epsilon \rho_\text{noisy},  
\end{align*}
for an arbitrary state $\rho_\text{noisy}$, so that $\varepsilon$ controls the model misspecification between the true physical prior and the assumed prior.



Given a set of POVM elements $\{\mat{A}_k\}$, the theoretical probabilities are given by  Born's rule, $p_k(\theta) = \text{Tr}(\mat{A}_k \rho(\theta))$. The multinomial likelihood function, representing the probability of observing the data given a specific model parameter, is
\begin{equation}
    P(\D \mid \theta) \propto \prod_{k} [p_k(\theta)]^{c_k}
\end{equation}
where $\D$ consists of the features $X$ which contain the frequencies $f_k$ and $c_k := f_k N$ are the counts. For simplicity, in this example we exclude the responses $Y$ from $\D$ since it is not straightforward to incorporate them.

To perform Bayesian inference, one uses a uniform prior $\pi(\theta) = 1$ over the interval $\theta \in [0, 1]$, and then Eq.~\eqref{eq:posterior} gives the posterior distribution which quantifies the model's updated belief about the true parameter after observing the data.
%
To then form a credible interval, 
we extract the appropriate percentiles from the posterior distribution of $\theta$ (which is easy since this is 1 dimensional), which is then mapped directly to the fidelity since it is a linear function of $\theta$ for our modeling choice.

In order to compare the performance of CP with Bayesian model, we sample $\rho_\text{phy}$ with $\epsilon$ drawn uniformly between $0$ and $1.0$ to construct a dataset of size $10000$ which is then split equally to form the training and calibration sets. We train and calibrate a NN once using these two sets. For equally spaced $\epsilon$ values between $0.0$ to $0.1$, we draw fresh samples for the test set and run CP using our trained model. We average our results over $1000$ runs for each value of $\epsilon$ which involves drawing a new test set every time.

The results are shown in Fig.~\ref{fig:bayes}. When the $\theta$ prior and the likelihood are fully known, that is, when $\epsilon = 0$, the Bayesian approach achieves the desired coverage of $90\%$ and a lower interval length. For $\epsilon > 0$, Bayes increasingly undercovers in terms of coverage and 
does not attempt to adjust the interval length in order to compensate.
In contrast, our CP method (using a NN prediction model) achieves target coverage for different levels of model misspecification with slight over-coverage for $\epsilon > 0$. Since we trained and calibrated our model only once, the estimated quantile $\hat q_{n_{2}}$ remains fixed and thus the average length remains constant, i.e., $2\hat q_{n_{2}}$.

One might argue that the Bayesian model fails simply because its likelihood function is misspecified, while the neural network with conformal framework succeeds because it is trained directly on the misspecified data. However, this is exactly the point. In experimental quantum computing, exact models of system noise are often mathematically intractable or entirely unknown. The strength of the conformal prediction framework is precisely its ability to bypass the need for explicit parametric modeling, leveraging empirical calibration data to provide rigorous safety guarantees even when the underlying physical processes are poorly understood. The same argument is validated by Fig.~\ref{fig:bayes}. 

\begin{figure*}[ht]
    \centering

    \begin{subfigure}[t]{\textwidth}
        \centering
        \includegraphics[
            width=0.45\textwidth,
            trim=0.2cm 0.2cm 0.2cm 0.2cm,
            clip
        ]{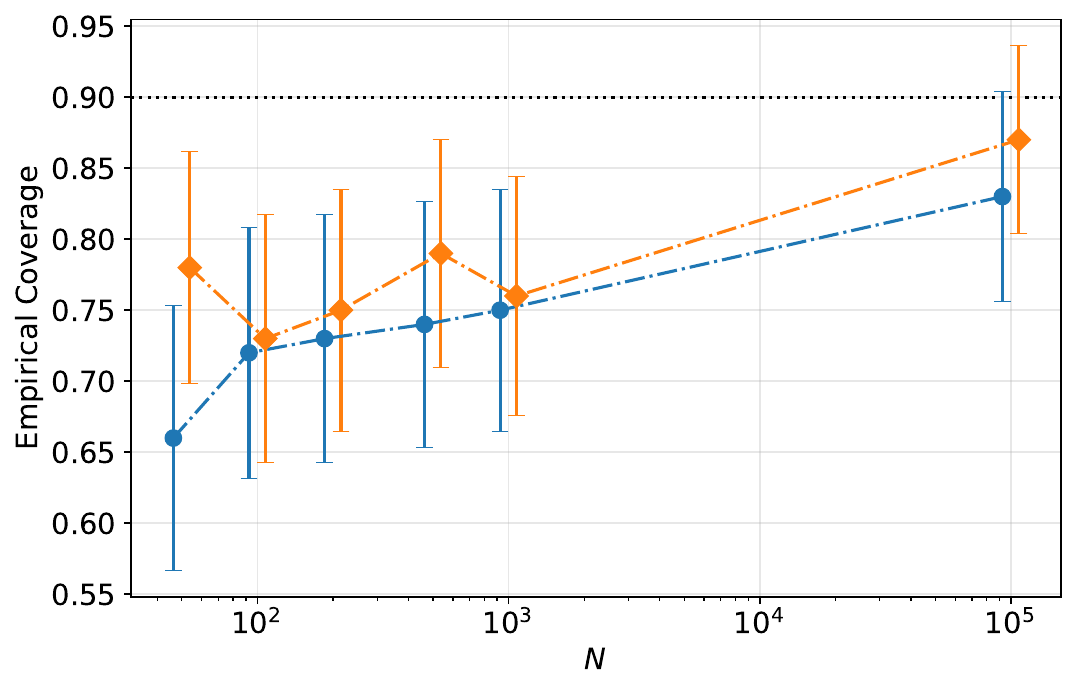}
        \hfill
        \includegraphics[
            width=0.45\textwidth,
            trim=0.2cm 0.2cm 0.2cm 0.2cm,
            clip
        ]{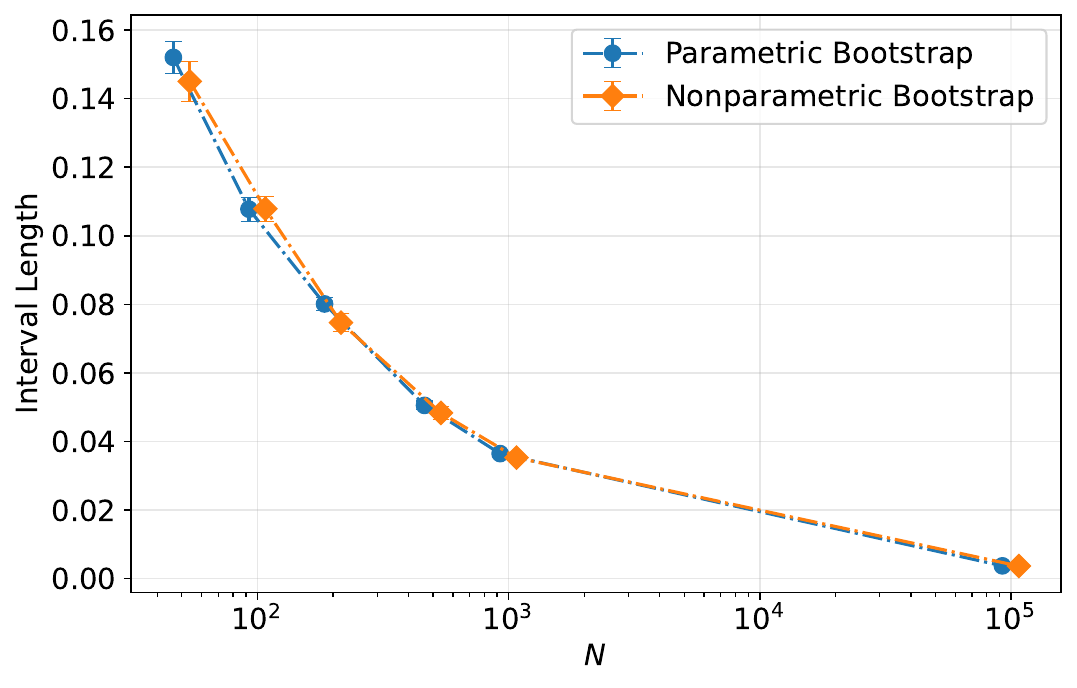}

    \end{subfigure}

    \caption{Impact of finite-sample effects on parametric and nonparametric bootstrap prediction 
    intervals under the non-asymptotic regime. For each bootstrap method, 
    empirical coverage (left) and interval length (right) are shown as a 
    function of shot count $N$ per POVM.}
    \label{fig:boot}
\end{figure*}
\subsection{The bootstrap in the non-asymptotic regime}
Confidence intervals generated via bootstrap resampling are not guaranteed to be correct. 
We computationally assess the asymptotic behavior of the parametric bootstrap for constructing prediction intervals by varying the number of measurement shots $N$ per POVM element. For a fixed underlying quantum state $\rho$, we first generate a single realization of measurement data by sampling outcomes from the multinomial distribution specified by Born’s rule with $N$ shots per POVM element. From this data, a state estimate $\hat{\rho}$ is obtained; we use MLE but PLS could be used as well.
Resampling is done by either redrawing from these outcomes with replacement (nonparametric bootstrap), or 
treating $\hat{\rho}$ as the parametric model and resampling measurement outcomes using the Born probabilities obtained with $\hat{\rho}$, again using $N$ shots per POVM element (parametric bootstrap). Each bootstrap instance is used to compute a new estimate of the state and the corresponding fidelity. This procedure yields an empirical bootstrap distribution from which prediction intervals are constructed using the percentile method.  

For each realization of the original measurement data, we generate $1000$ such bootstrap samples, requiring $1000$ MLE solves. To obtain error bars, we repeat the entire procedure over $100$ independent realizations of the measurement data and average the results, leading to $100{,}000$ MLE solves for entire simulation.

This procedure is repeated for increasing values of $N$, allowing us to probe the transition from non-asymptotic to asymptotic regimes. In this context, the asymptotic regime corresponds to $N$ $\to \infty$, under which the multinomial sampling noise diminishes, the estimator $\hat{\rho}$ becomes
more accurate due to the consistency properties of the MLE,
 and the bootstrap distribution converges to the true sampling distribution induced by $\hat{\rho}$. As shown in Fig.~\ref{fig:boot}, the limitations of the bootstrap framework in non-asymptotic regimes are readily apparent. At lower shot count, both bootstrap methods exhibit significant undercoverage alongside excessively wide interval lengths. As $N$ becomes practically large, the empirical coverage starts to converge to the target level while the corresponding interval length systematically reduces.

\section{Validity and efficiency of conformal prediction for fidelity estimation} \label{sec:results}

\begin{figure*}[htb]
    \centering

    \hspace{.2cm}\includegraphics[width=.69\textwidth, trim=0.2cm 0.1cm 0.2cm 0.2cm, clip]{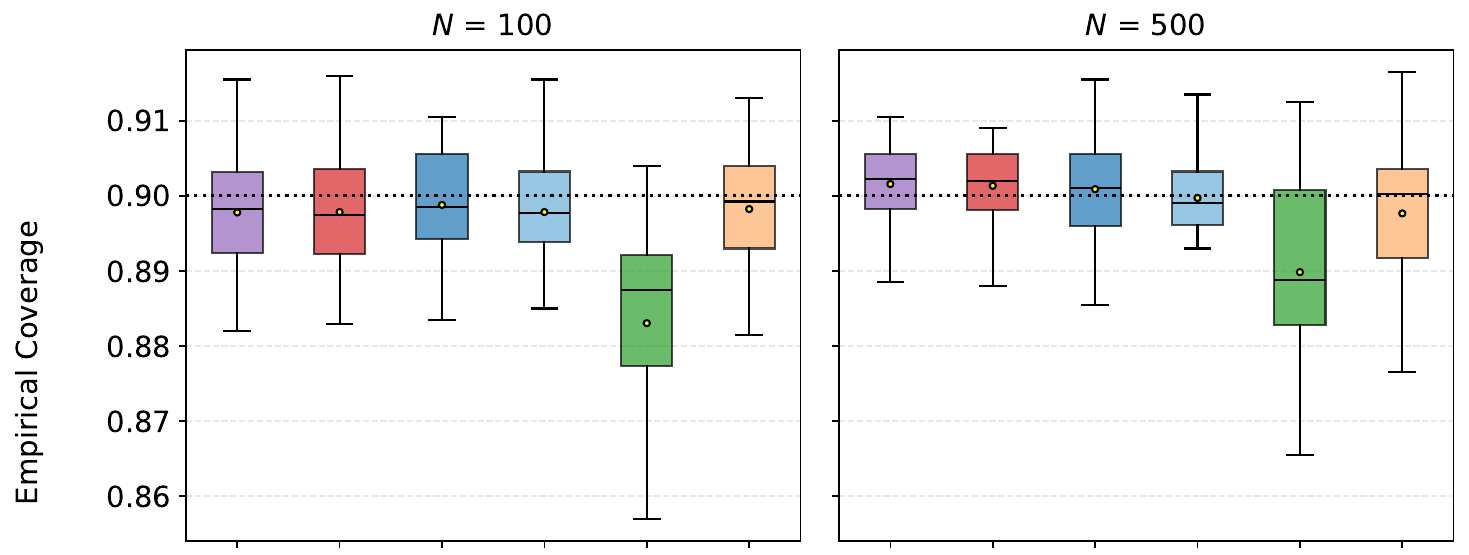}
    \includegraphics[width=.7\textwidth, trim=0.2cm 0.2cm 0.2cm 0cm, clip]{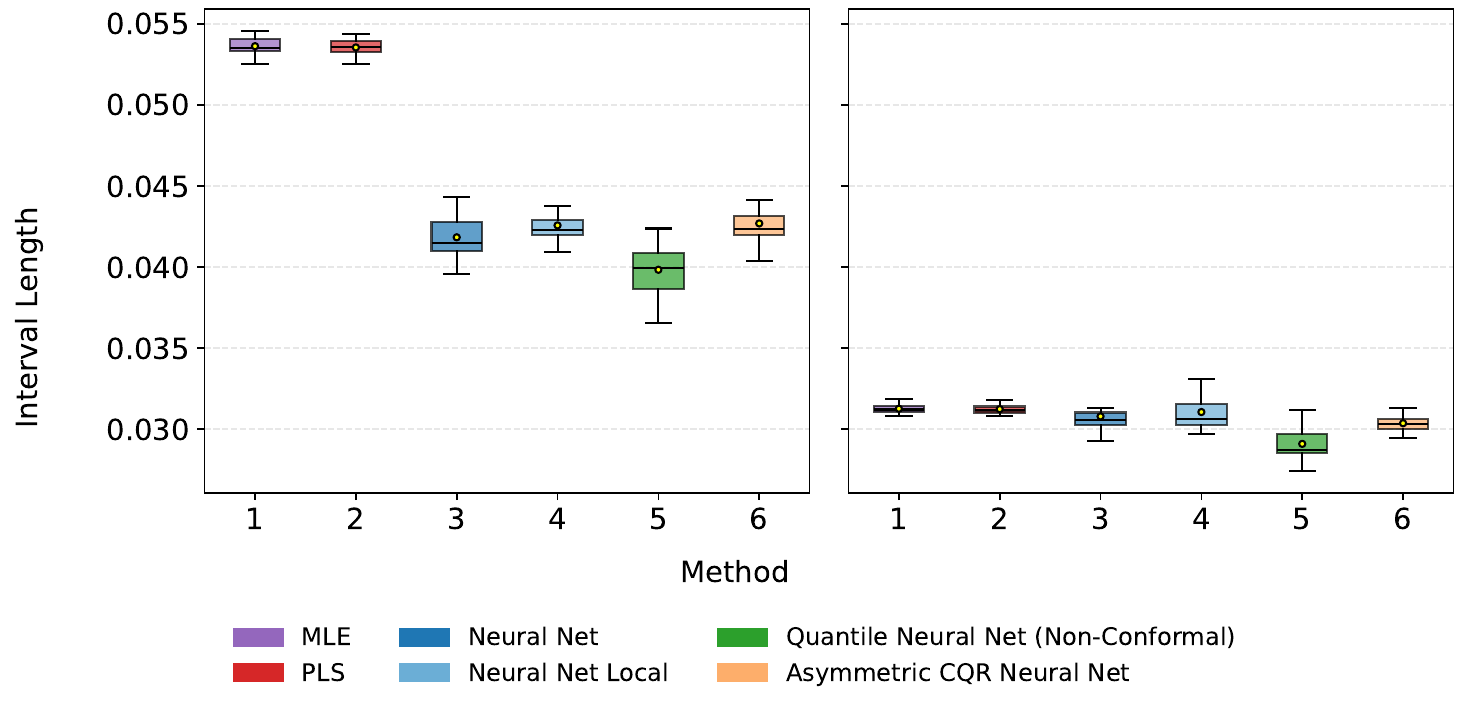}

    \caption{Statistical performance of conformalized fidelity estimators. 
    (Top) Empirical coverage and (Bottom) 
    interval length as a function of shot count $N$ for $q=4$ qubits. In the box plots, the yellow dots represent the mean and the solid horizontal black lines represent the median.}
    \label{fig:cov}
\end{figure*}

\subsection{Proposed framework for prediction intervals}

We evaluate several variants of our framework described in Algo.~\ref{alg:cap}, as well as a non-CP baseline:
\begin{description}
    \item[\texttt{MLE}] Estimates fidelity using the MLE method of \S\ref{subsec:plugin}, and does our UQ framework Algo.~\ref{alg:cap} using standard conformal prediction.
    \item[\texttt{PLS}] Estimates fidelity using the PLS method of \S\ref{subsec:plugin}, and does our UQ framework Algo.~\ref{alg:cap} using standard conformal prediction.
    \item[\texttt{Neural Net}] Estimates fidelity using the NN method of \S\ref{subsec:NN}, and does our UQ framework Algo.~\ref{alg:cap} using standard conformal prediction.
    \item[\texttt{Neural Net Local}] Same as \texttt{Neural Net} but uses the adaptive CP variant described in \S\ref{subsubsec:adaptivesplit}
    \item[\texttt{Quantile Neural Net (Non-Conformal)}] Estimates lower and upper fidelity quantiles directly using a two-output neural network with the same architecture as \texttt{Neural Net}, trained by minimizing the pinball loss; see \S\ref{subsubsec:quantilesplit} (uses $\D_1$ only).

    \item[\texttt{Asymmetric CQR Neural Net}] Same as \texttt{Quantile Neural Net (Non-Conformal)}, but applies conformal calibration to the lower and upper quantile estimates (using $\D_2$) as described in \S\ref{subsubsec:quantilesplit}.
    
\end{description}

In this section, we consider the traditional prediction interval, described by Eq.~\eqref{eq:actualGuarantee}. 
We assess the coverage and the interval lengths of these methods as functions of both number of qubits and shot count per POVM $N$. We generate a dataset of Hilbert-Schmidt states which are full-rank. We further split this dataset into train, calibration and test which ensures performance evaluation under exchangeability. Specifically, we first split the dataset of size 20000 into \(80\%\) training data and \(20\%\) test data. The training portion is then divided equally into two disjoint subsets, yielding proper training and calibration data.
To ensure statistical stability, all performance metrics are reported over 20 independent trials for each configuration of qubit and shot count which involves random splitting of the dataset into three sets as mentioned above. A baseline architecture is shared across all NN-based frameworks and it is the one described in \S\ref{subsec:models}. 

As shown in Fig.~\ref{fig:cov} for $q=4$ qubits (Figs.~\ref{fig:cov-all} and \ref{fig:len-all} in the appendix for $2$ and $6$ qubits), all conformalized methods successfully maintain the target coverage level of $1-\alpha=0.9$, independent of the number of shots $N$, as expected by the theory. In contrast, the non-conformal baseline consistently exhibits undercoverage. 

Regarding efficiency, the NN-based CP methods yield comparable interval lengths. While the non-conformal method produces shorter intervals, this is due to over-coverage. Notably, NN-based estimators significantly outperform classical methods in the low-shot regime.  Increasing the shot count from $N=100$ to $N=500$, all the methods have tighter intervals, as one would expect. In this higher shot count regime, 
classical estimators become competitive with the NN estimators as noise in the measurements diminishes. Due to the prohibitive computational cost of classical reconstruction (MLE and PLS) in higher dimensions, comparisons for $6$-qubit system are restricted to NN-based approaches.

The observed superiority of NN-based intervals can be attributed to the relationship between interval length and estimator variance. Traditional PLS and MLE estimators operate without explicit regularization, making them highly sensitive to statistical noise in the measurement data. Conversely, the neural network—having been trained on measurements from states sampled from an appropriate prior—implicitly adopts this distribution as an inductive bias, leading to more concentrated residuals and therefore, allowing the conformal framework to produce tighter bounds.

\begin{figure*}[hbt]
    \centering

    \begin{subfigure}[t]{\textwidth}
    \centering
        \includegraphics[width=0.45\textwidth, trim=0.2cm 0.2cm 0.2cm 0.2cm, clip]{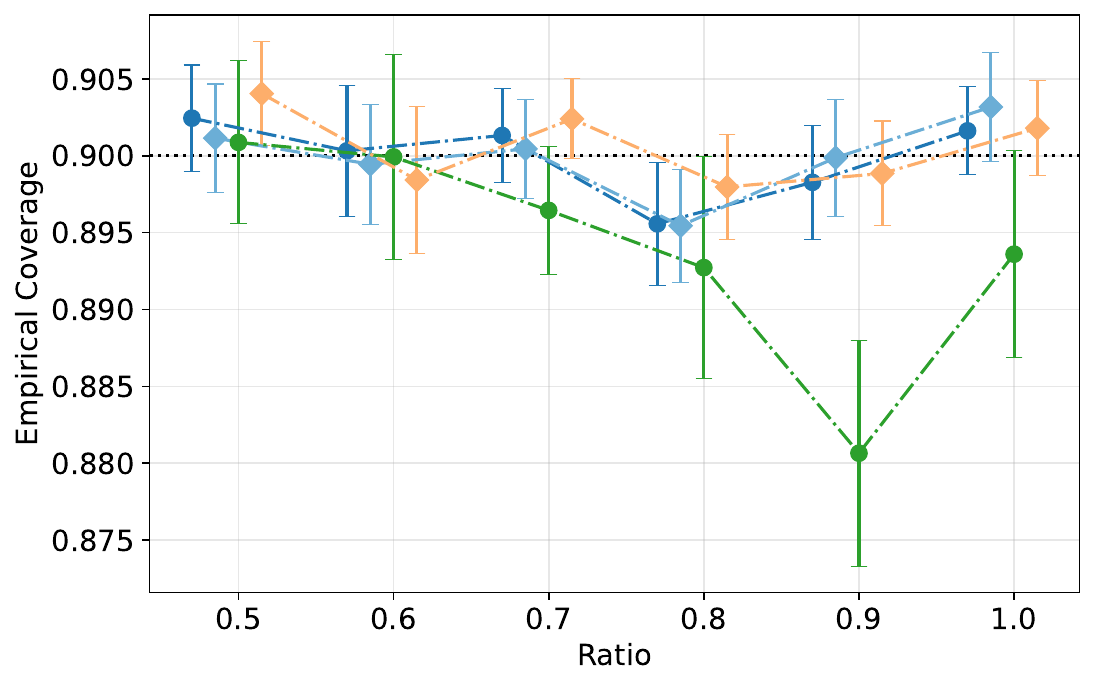}
        \label{fig:incomplete-cov}
        \hfill
        \includegraphics[width=0.45\textwidth, trim=0.2cm 0.2cm 0.2cm 0.2cm, clip]{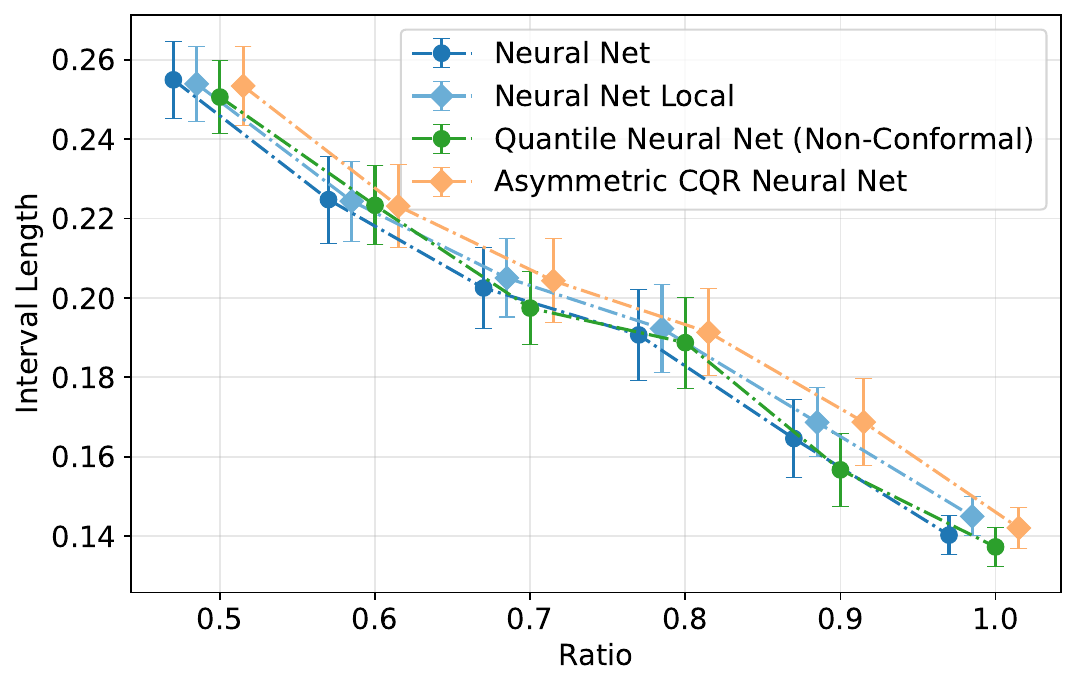}
        \label{fig:incomplete-len}
    \end{subfigure}

    \caption{Empirical coverage (left) and interval length (right), as a function of the fraction of the total POVM elements used for measurement for $q=2$ qubits. $x$-locations are jittered slightly for presentation purposes.}
    \label{fig:incomplete}
\end{figure*}
\subsection{Conformal prediction under incomplete measurements} To evaluate the robustness of conformal prediction under  tomographically incomplete measurement schemes, 
we randomly subsample the full set of  $d^2$ Pauli POVMs and retain only a fraction of the measurements, varied from $0.5$ and $1.0$. 
As illustrated in Fig.~\ref{fig:incomplete}, the conformal-based methods roughly achieve the target coverage level of $90\%$ independent of the POVM subsampling, but the interval lengths are much wider in the undersampled regime, reflecting the lower information.
%
For instance, the average interval length for the standard NN expands from approximately $0.14$ under complete measurements to over $0.25$ in the highly incomplete regime. This widening reflects the physical ambiguity of incomplete quantum state tomography because multiple distinct states can generate identical statistics for a restricted measurement set. The conformal framework  dynamically widens its intervals to safely bound the true fidelity.

\subsection{Conditional coverage with conformal prediction}\label{subsec:cond-cov}
The results obtained earlier demonstrate the successful application of the conformal framework in state fidelity estimation, guaranteeing marginal coverage (Eq.~\eqref{eq:actualGuarantee}) under the exchangeability of calibration and test data. Going further, we are also interested in examining the validity and efficiency of prediction intervals conditional on the specific quantum state $\rho$, as defined in Eq.~\eqref{eq:state-conditional}. Since exact state-conditional coverage is mathematically intractable to guarantee with finite data, we approximate this local behavior by analyzing the empirical coverage across uniform fidelity bins. This partitioning allows us to evaluate whether our intervals maintain localized reliability across different regions of the Hilbert space, serving as a physically meaningful proxy for true state-conditioned validity. 


Consider the following distribution of states that is a convex mixture of a target pure state $\sigma$ and a stochastic noise component $\rho^{\prime}$
\begin{equation}\label{state-distribution}
     \rho=
    \eta \sigma + (1-\eta)\rho^{\prime}
\end{equation}
To construct the training and calibration datasets used in the conditional coverage analysis of the following subsections, we let $\rho^{\prime}$ be sampled according to the Hilbert--Schmidt measure so that 
$
    \rho_{\mathrm{train}} = \eta \sigma + (1-\eta)\rho_{\mathrm{HS}}.
$
We first sample a target fidelity $F_{\mathrm{tar}}$ uniformly from $[\delta,1]$, for small $\delta$. For any $\rho_{\mathrm{HS}}$, let
$
    c = F(\sigma,\rho_{\mathrm{HS}}).
$
Since $\sigma$ is pure
\begin{equation*}
    F(\sigma,\rho_{\mathrm{train}})
    =
    \eta + (1-\eta)c.
\end{equation*}
Therefore, choosing
\begin{equation*}
    \eta
    =
    \frac{F_{\mathrm{tar}}-c}{1-c} \le 1
\end{equation*}
ensures that
$F(\sigma,\rho_{\mathrm{train}})=F_{\mathrm{tar}}$.
Only realizations for which $\eta>0$ are retained, which is almost always satisfied since $c$ is typically near zero, ensuring that $\rho_{\mathrm{train}}$ is a valid convex mixture.

We implement a $Y$-conditioned conformal prediction framework mentioned in \S\ref{subsubsec:conditional} to ensure statistical validity across the full range of fidelity values. The fidelity domain $[0,1]$ is partitioned into 10 uniform bins via a discretization function. While the underlying neural network architecture remains unchanged, we compute bin-specific quantiles by evaluating the conformity scores of the calibration data residing within each respective bin. During inference, the point estimate produced by the network determines the corresponding bin assignment for each test instance and the associated bin-specific quantile is then retrieved to construct the prediction interval. The empirical coverage within each fidelity bin is then computed as the fraction of test instances whose true fidelity lies within the corresponding prediction interval.

\subsubsection{Conditional coverage under Born's rule}
\begin{figure*}[ht]
    \centering

    \begin{subfigure}[t]{\textwidth}
    \centering
        \includegraphics[width=0.45\textwidth, trim=0.2cm 0.2cm 0.2cm 0.2cm, clip]{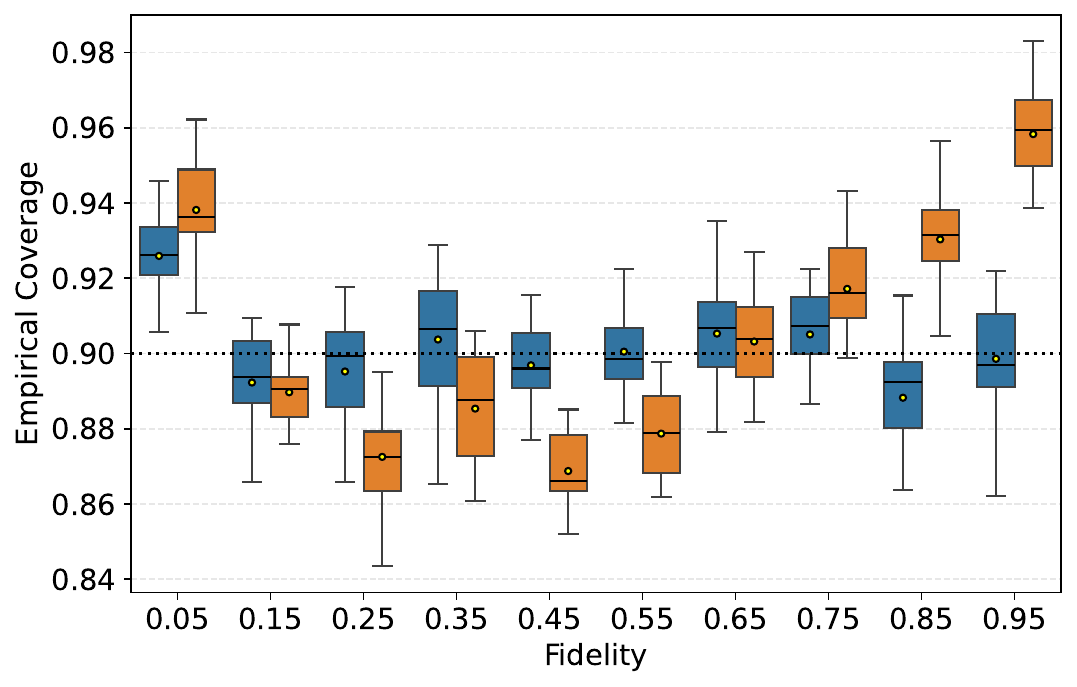}
        \label{fig:TBDA}
    \hfill
        \includegraphics[width=0.45\textwidth, trim=0.2cm 0.2cm 0.2cm 0.2cm, clip]{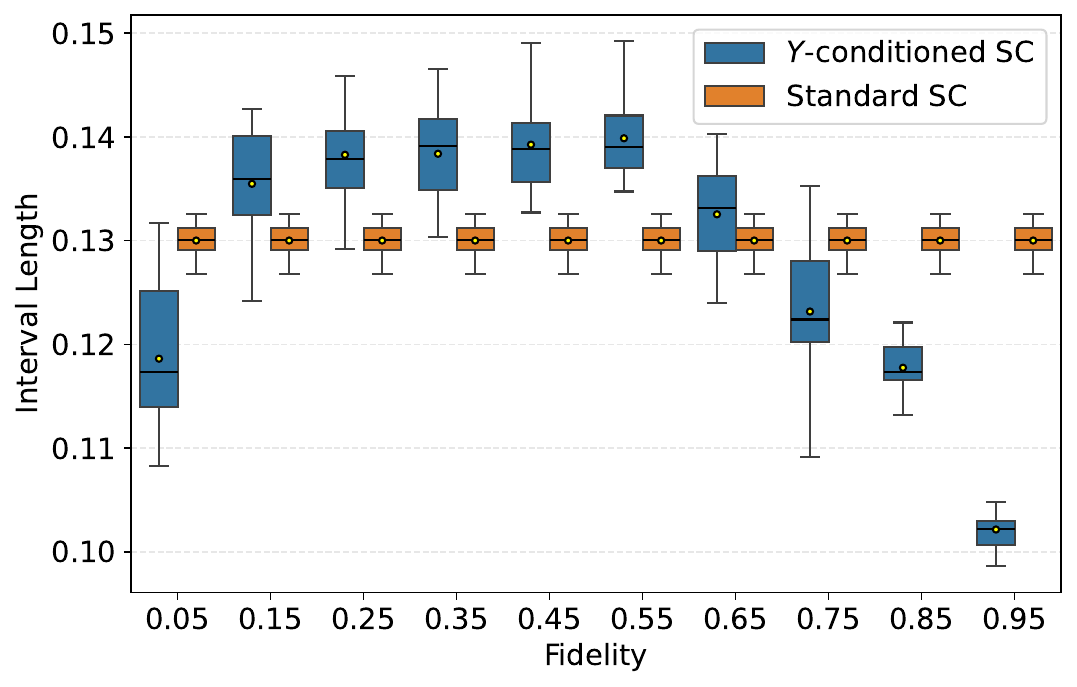}
        \label{fig:TBDB}
    \end{subfigure} 
    
    \caption{Conditional coverage analysis across fidelity regimes for $q=2$ qubits, averaging over Born's rule. Empirical coverage is plotted as a function of state fidelity, partitioned into $10$ equal-width bins, where $x$-axis ticks are mid-points of each bin. Results are averaged over $20$ independent splits of the training and calibration data.}
    \label{fig:cond-cov-born}
\end{figure*}
We evaluate the conditional coverage properties of the prediction intervals in the presence of finite-shot measurement noise (Born's rule). To this end, we sample test states from the same prior ensemble \eqref{state-distribution} as the training and calibration data, specifically stratifying the selection to ensure a uniform representation of true fidelities across the domain $[0,1]$ relative to the reference state $\sigma$. 

For each of the $100$ sampled fidelity values, we generate a test state and use this test state to obtain $100$ independent measurement realizations. Thus, each fidelity value contributes $100$ test samples. This allows us to assess coverage conditional on stratified fidelity regions while averaging over the randomness introduced by finite-shot measurement noise. Because the empirical coverage is averaged over repeated experiments for a fixed $\rho$, this experiment is designed to test the validity of Eq.~\ref{eq:state-conditional} which is the desired property (but not quite the same as the guaranteed property of Eq.~\eqref{eq:actualGuarantee}).
The resulting metrics are averaged over $20$ independent splits of the training and calibration data to assess variability across runs.

As illustrated in Fig.~\ref{fig:cond-cov-born}, standard split conformal prediction fails to provide uniform local reliability, exhibiting severe under- and over-coverage when the test instances are binned by their true fidelity. Conversely, the $Y$-conditioned split conformal method successfully maintains the nominal target coverage across the majority of the fidelity bins. This demonstrates that conditioning the calibration process on the target space yields robust local adaptivity, protecting the experimentalist against overconfident or overly conservative intervals depending on the quality of the prepared quantum state.

\subsubsection{Conditional coverage under Born's rule and out-of-distribution (OOD) noise} 

To evaluate the robustness of the conformal prediction procedure under a distribution shift in the state-preparation process, we generate the test ensemble using a depolarizing noise model. Specifically, we set $\rho^{\prime} = I/d$ in Eq.~\eqref{state-distribution}, yielding
$
    \rho_{\mathrm{test}} = \eta \sigma + (1-\eta)\frac{I}{d}.
$
The training and calibration ensembles are generated as before according to $\rho_{\mathrm{train}}$. For each fidelity value, a test state is generated according to $\rho_{\text{test}}$ which is, in turn, used to generate $100$ independent measurement realizations. The resulting metrics are averaged over $20$ independent splits of the training and calibration data.

\begin{figure*}[ht]
    \centering
    
    \begin{subfigure}[t]{\textwidth}
    \centering
        \includegraphics[width=0.45\textwidth, trim=0.2cm 0.2cm 0.2cm 0.2cm, clip]{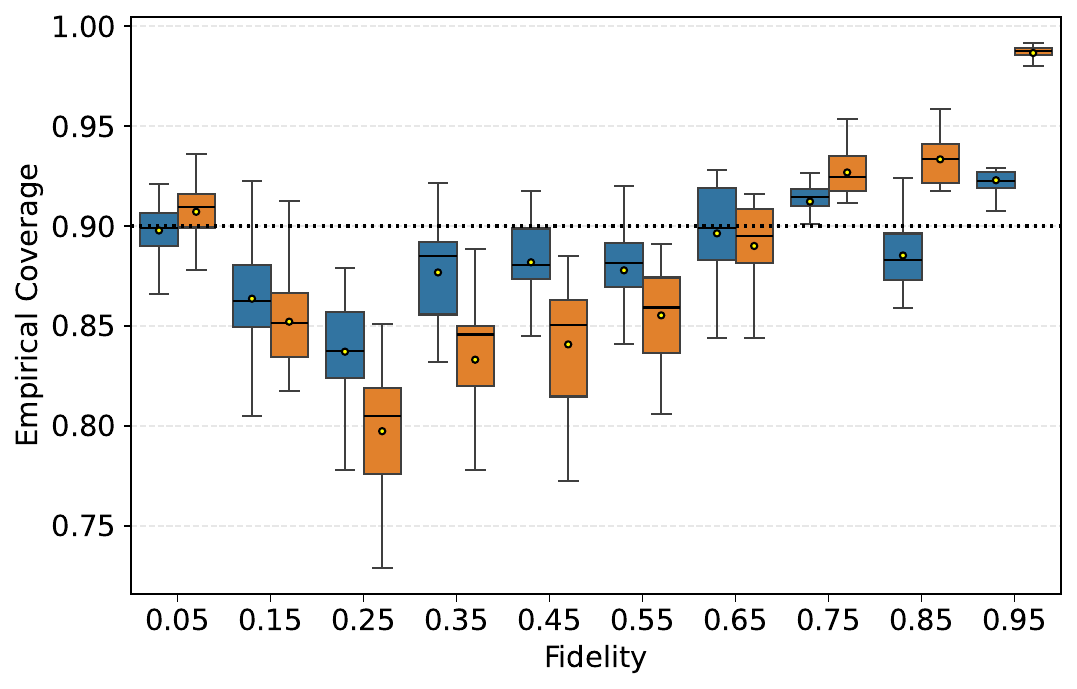}
      \hfill
        \includegraphics[width=0.45\textwidth, trim=0.2cm 0.2cm 0.2cm 0.2cm, clip]{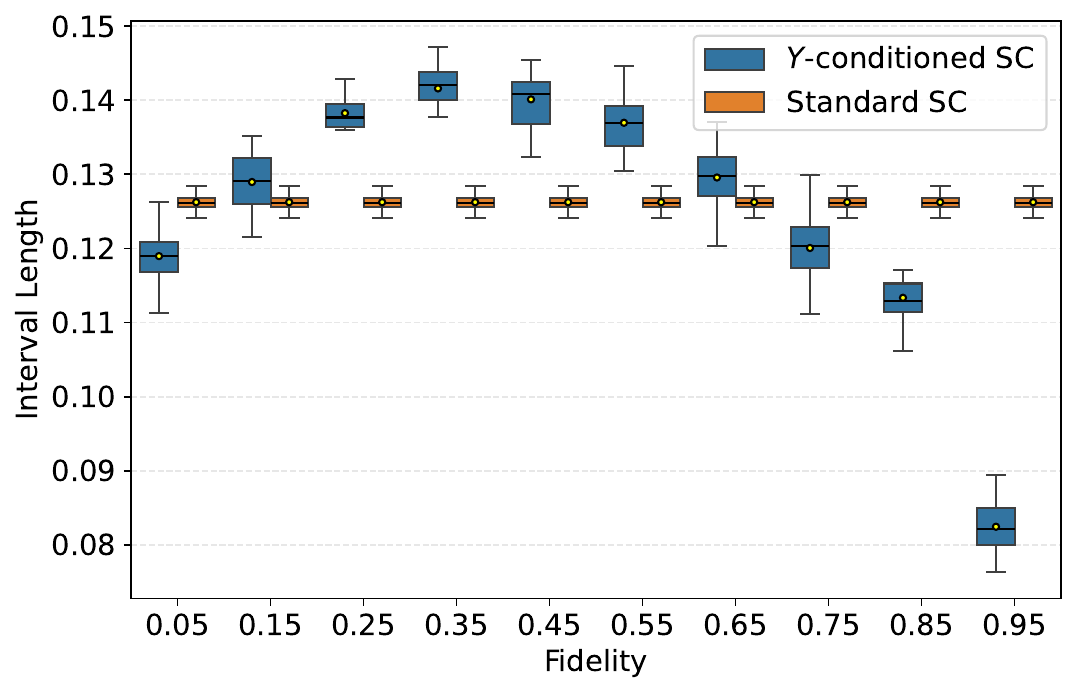}
    \end{subfigure}
    
    \caption{Conditional coverage analysis across fidelity regimes for $q=2$ qubits in case of out-of-distribution depolarizing noise in test data. Empirical coverage as a function of state fidelity, partitioned into $10$ equal-width bins, where $x$-axis ticks are mid-points of each bin. Results are averaged over $20$ independent splits of the training and calibration data.}
    \label{fig:conditional_n=2}
\end{figure*}

As shown in Fig.~\ref{fig:conditional_n=2}, both methods exhibit coverage degradation in the low-fidelity regime ($\eta \to 0$), where the state approaches the maximally mixed state $\frac{1}{d}\identity$ and the measurement statistics become isotropic. However, while the standard split conformal method fails to recover across the domain, the $Y$-conditioned split conformal method demonstrates remarkable resilience in the mid-to-high fidelity regimes 
$(\eta \gtrsim 0.5$).
By conditioning on the target space, the method effectively adapts to the OOD noise profile, successfully restoring and maintaining the nominal target coverage where the standard CP approach fails.
    
\section{Conclusion}\label{sec:conclusion}

In this work, we have addressed a critical challenge in near-term quantum characterization: providing reliable, computationally efficient uncertainty quantification for quantum state fidelity estimation under different regimes. While traditional approaches like bootstrap resampling suffer from severe undercoverage and inflated interval widths in non-asymptotic regimes, and Bayesian inference remains highly vulnerable to model misspecification, conformal prediction offers a powerful, distribution-free framework to guarantee statistical coverage. 

Our core contribution was to address the gap between the marginal guarantees provided by standard conformal prediction and the state-conditional coverage demanded by experimentalists. By implementing a hybrid strategy that wraps Bayesian priors—such as the Hilbert-Schmidt measure—within a split conformal framework, we demonstrate that the underlying neural network acquires an inductive bias reflecting the global geometry of the quantum state space. This geometric bias allows the estimator to maintain tight, highly informative prediction intervals even when subjected to severe, out-of-distribution noise channels, such as depolarizing noise.

Furthermore, our systematic evaluation highlights that while standard split conformal methods fail to provide uniform reliability across different state configurations, the $Y$-conditioned split conformal method is better at achieving approximate state-conditional validity. By conditioning the calibration architecture directly on the target fidelity space, this approach dynamically adapts to the local structure of the Hilbert space, ensuring robust coverage for specific physical state realizations. Ultimately, this framework provides experimentalists with a scalable, model-agnostic, and mathematically rigorous ``safety wrapper" for quantum hardware validation, rendering fast machine-learning-based tomography practically viable for the certification of NISQ-era devices.

\begin{acknowledgments}
This material is based upon work supported by the National
Science Foundation under Grant No.\ 2106834.
\end{acknowledgments}

\appendix

\section{Neural Net training}\label{appendix}

\begin{figure*}[ht]
    \centering

    \begin{subfigure}[t]{0.4\textwidth}
        \includegraphics[width=\textwidth, trim=0.2cm 0.2cm 0.2cm 0.2cm, clip]{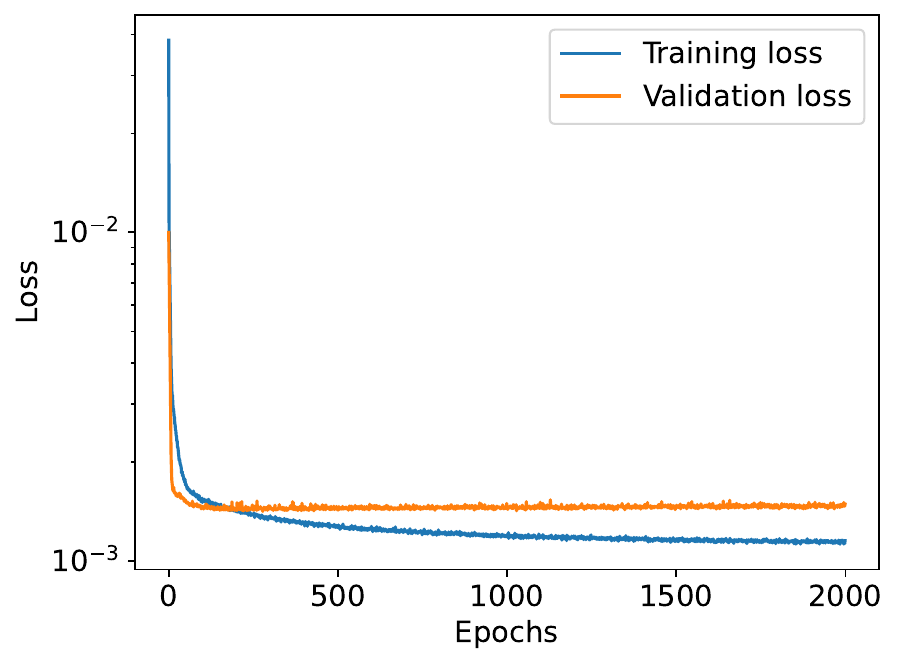}
        \caption{}
    \end{subfigure}
    \hfill
    \begin{subfigure}[t]{0.4\textwidth}
        \includegraphics[width=\textwidth, trim=0.2cm 0.2cm 0.2cm 0.2cm, clip]{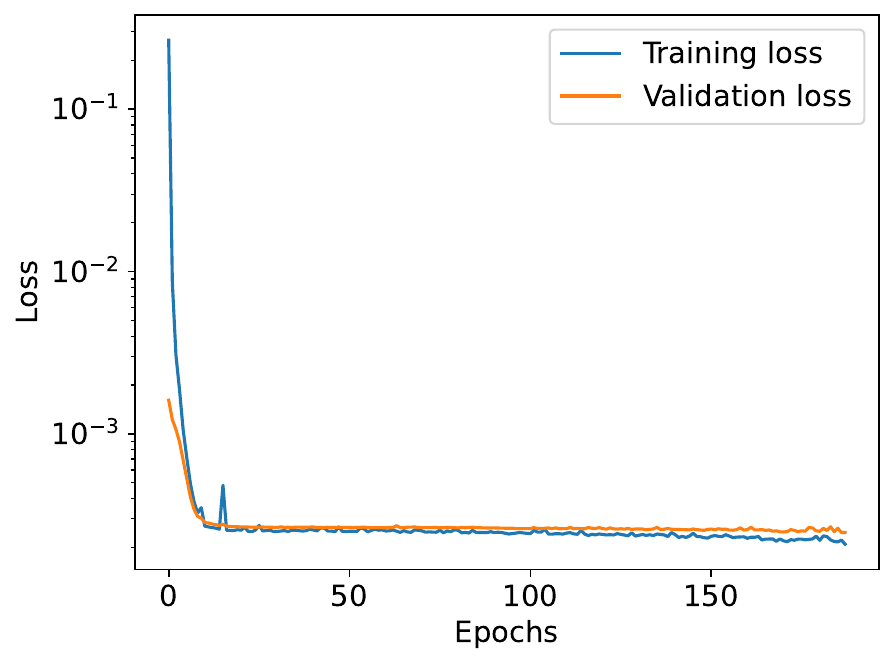}
        \caption{}
    \end{subfigure}
    
    \caption{Training and validation loss trajectories. The mean squared error (MSE) is plotted as a function of training epochs for (a) $q=2$ and (b) $q=6$ qubits. In both cases, the validation loss 
    is sufficiently low --- there is no need for extreme accuracy since the inherent uncertainty in the estimate is typically $10^{-2}$ or larger.
    }
    \label{fig:nn-loss}
\end{figure*}

Fig.~\ref{fig:nn-loss} shows the training and validation curves for training the neural network. For both $2$ and $6$ qubits, the validation loss 
    is sufficiently low --- there is no need for extreme accuracy since the inherent uncertainty in the estimate is typically $10^{-2}$ or larger.

\section{Additional plots}

\begin{figure*}[ht]
    \centering

    \begin{subfigure}[t]{\textwidth}
        \centering
        \includegraphics[width=\textwidth, trim=0.2cm 0.2cm 0.2cm 0.2cm, clip]{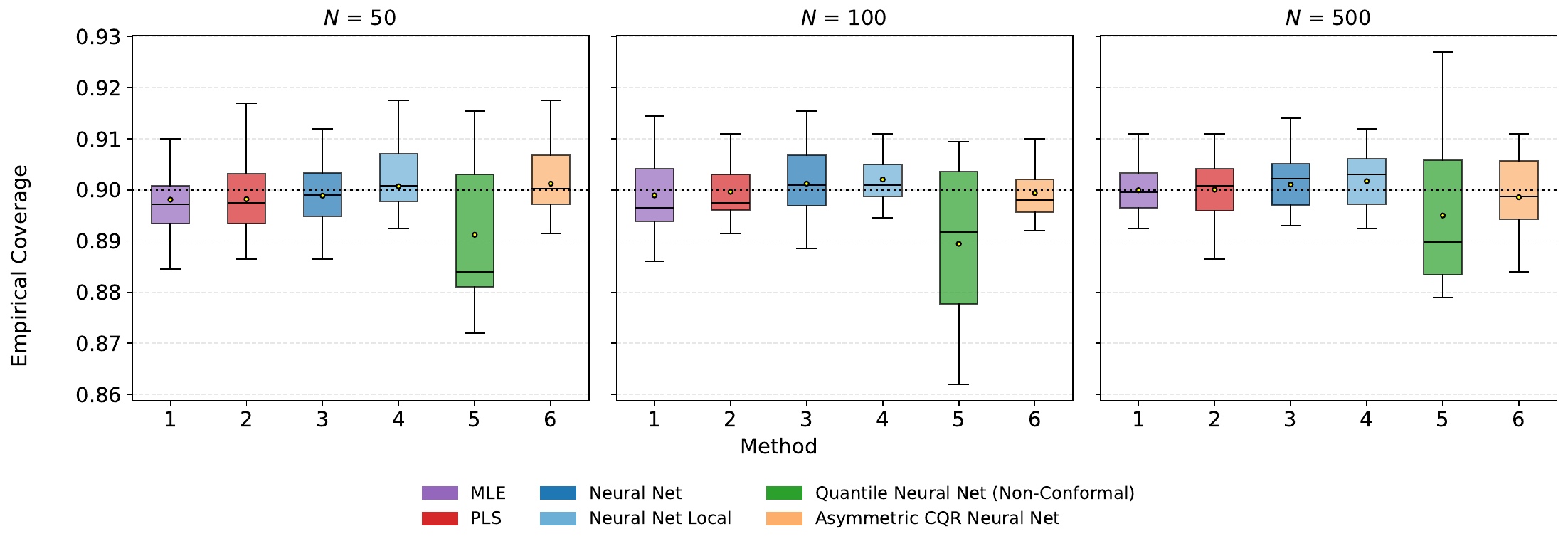}
        \caption{Coverage for 2 qubits}
        \label{fig:cov-2}
    \end{subfigure}
    \vspace{0.5cm}


    \begin{subfigure}[t]{\textwidth}
        \centering
        \includegraphics[width=\textwidth, trim=0.2cm 0.2cm 0.2cm 0.2cm, clip]{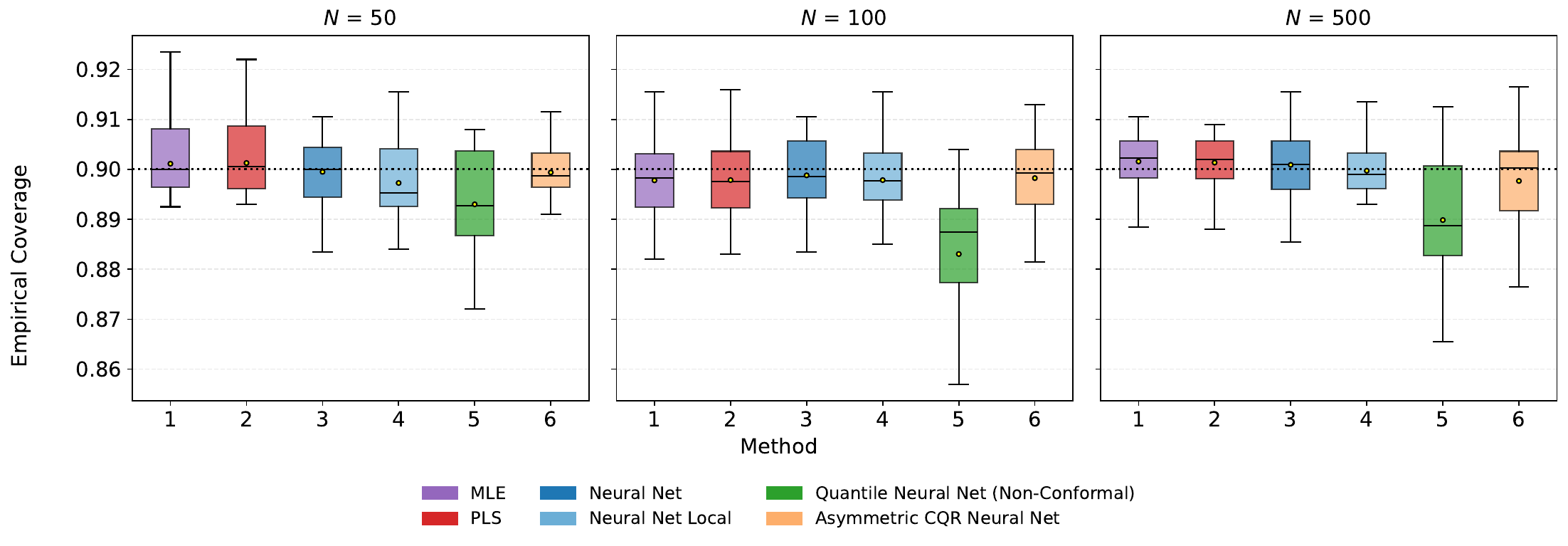}
        \caption{Coverage for 4 qubits}
        \label{fig:cov-4}
    \end{subfigure}
    \vspace{0.5cm}

    \begin{subfigure}[t]{\textwidth}
        \centering
        \includegraphics[width=\textwidth, trim=0.2cm 0.2cm 0.2cm 0.2cm, clip]{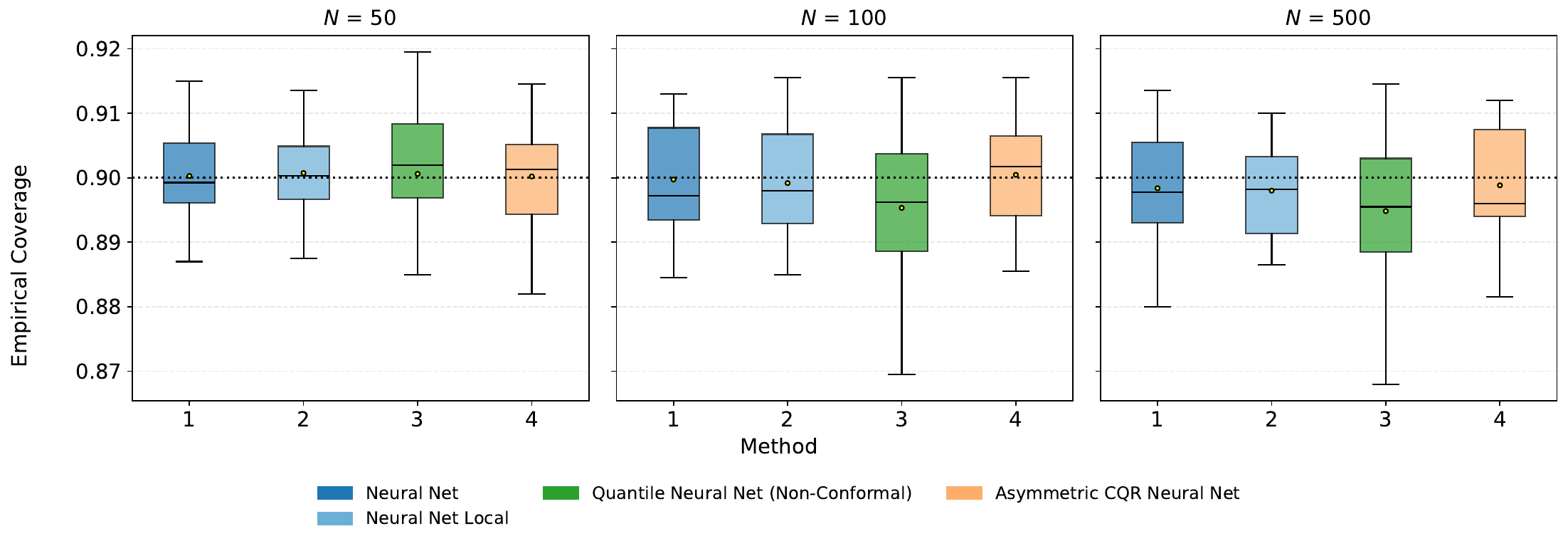}
        \caption{Coverage for 6 qubits}
        \label{fig:cov-6}
    \end{subfigure}

    \caption{Statistical performance of conformalized fidelity estimators. Empirical coverage for (a) $q=2$ qubits, (b) $q=4$ qubits and (c) $q=6$ qubits as a function of shot count $N$. In the box plots, the yellow dots represent the mean and the solid horizontal black lines represent the median. MLE and PLS were not run for 6 qubits due to their computational cost.
    }
    \label{fig:cov-all}
\end{figure*}

\begin{figure*}[ht]
    \centering

    \begin{subfigure}[t]{\textwidth}
        \centering
        \includegraphics[width=\textwidth, trim=0.2cm 0.2cm 0.2cm 0.2cm, clip]{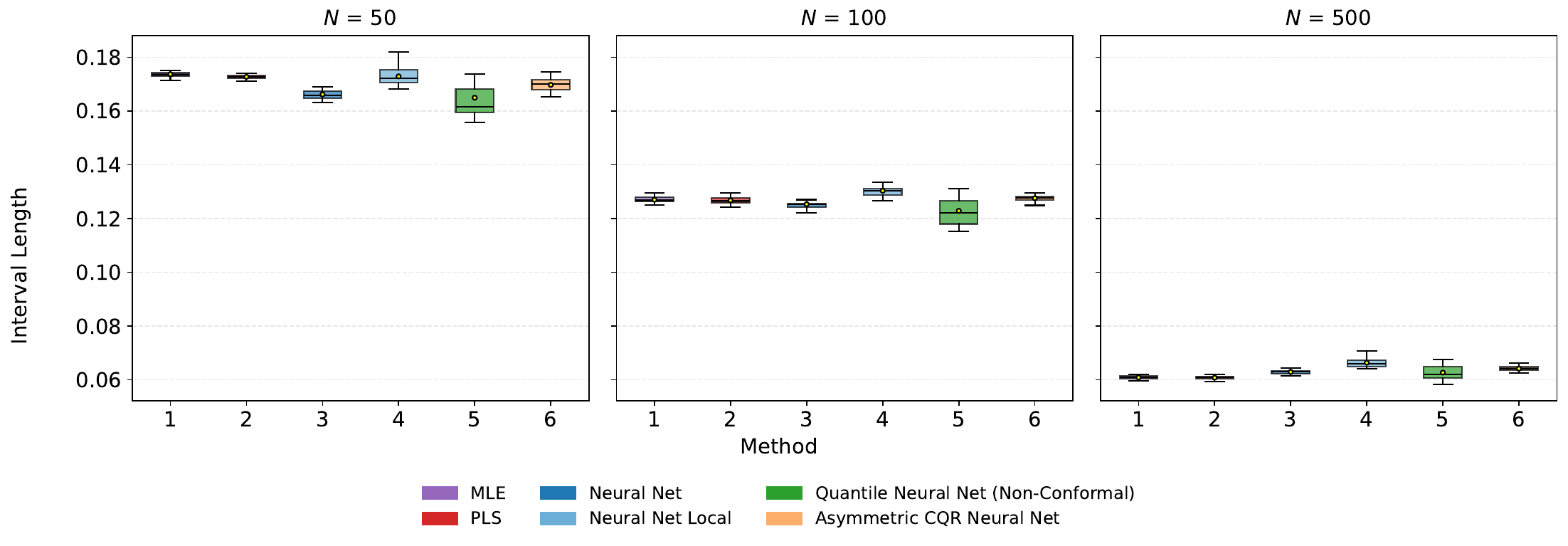}
        \caption{Interval Length for 2 qubits.}
    \end{subfigure}
    \vspace{0.5cm}


    \begin{subfigure}[t]{\textwidth}
        \centering
        \includegraphics[width=\textwidth, trim=0.2cm 0.2cm 0.2cm 0.2cm, clip]{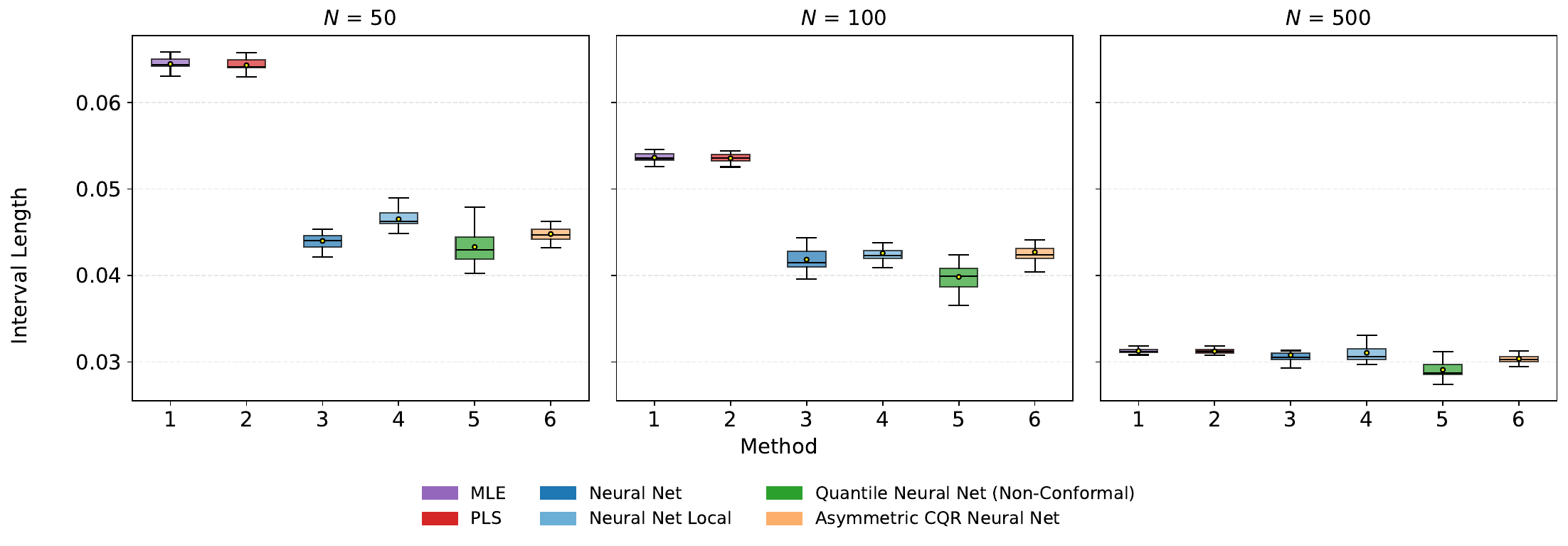}
        \caption{Interval Length for 4 qubits.}
    \end{subfigure}

    \vspace{0.5cm}

    \begin{subfigure}[t]{\textwidth}
        \centering
        \includegraphics[width=\textwidth, trim=0.2cm 0.2cm 0.2cm 0.2cm, clip]{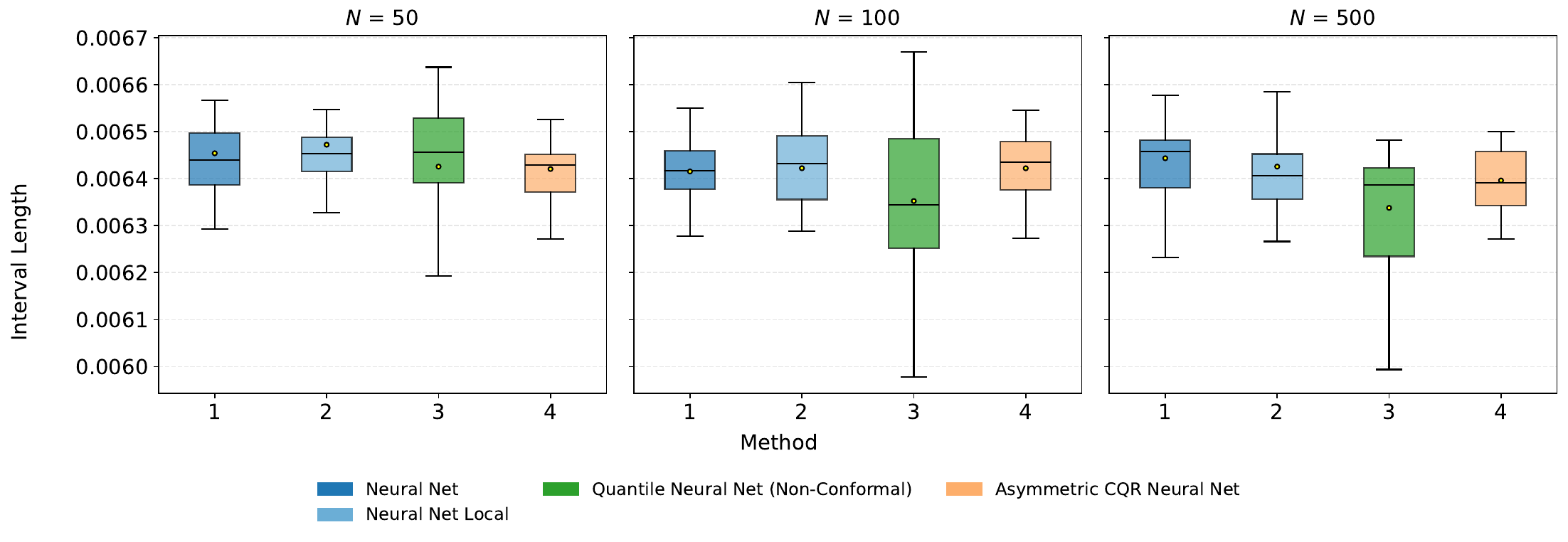}
        \caption{Interval Length for 6 qubits.}
    \end{subfigure}

    \caption{Statistical performance of conformalized fidelity estimators. Interval length for (a) $q=2$, (b) $q=4$ and (c) $q=6$ as a function of shot count $N$. In the box plots, the yellow dots represent the mean and the solid horizontal black lines represent the median.}
    \label{fig:len-all}
\end{figure*}

Figs.~\ref{fig:cov-all} and \ref{fig:len-all} supplement Fig.~\ref{fig:cov} from the main text by adding the $q=2$ and $q=6$ qubit cases.
The results are mostly qualitatively similar, though we note that the non-CP method improves its coverage level at $6$ qubits.

\bibliography{references, ref}

\end{document}